\documentclass{aa}  

\usepackage{graphicx}
\usepackage[varg]{txfonts}
\usepackage{natbib}
\usepackage{hyperref}

\bibpunct{(}{)}{;}{a}{}{,} 

\usepackage{color}
\begin{document}
%
\title{Three-dimensional mixing and light curves: constraints on the progenitor
       of supernova 1987A\thanks{%
   Data of the presupernova models for blue supergiants, the angle-averaged
      profiles of the 3D explosion models, and the corresponding bolometric
      light curves are available in electronic form at the CDS via
      anonymous ftp to cdsarc.u-strasbg.fr (130.79.128.5) or via
      \url{http://cdsweb.u-strasbg.fr/cgi-bin/qcat?J/A+A/}}
       }

\author{V.~P.~Utrobin\inst{1,2} \and A.~Wongwathanarat\inst{1} \and
        H.-Th.~Janka\inst{1} \and E.~M\"uller\inst{1} \and T.~Ertl\inst{1} \and
        S.~E.~Woosley\inst{3}}

\institute{
   Max-Planck-Institut f\"ur Astrophysik,
   Karl-Schwarzschild-Str. 1, 85748 Garching, Germany
\and
   State Scientific Center of the Russian Federation --
   Institute for Theoretical and Experimental Physics of
   National Research Center ``Kurchatov Institute'',
   B.~Cheremushkinskaya St. 25, 117218 Moscow, Russia
\and
   Department of Astronomy and Astrophysics, University of California,
   Santa Cruz, CA 95064, USA
}
\date{Received 24 December 2018 / Accepted 4 March 2019}
%
\abstract{%
With the same method as used previously,
   we investigate neutrino-driven explosions of a larger sample
   of blue supergiant models.
The blue supergiants were evolved as single-star progenitors.
The larger sample includes three new presupernova stars.
The results are compared with light-curve
   observations of the peculiar type IIP supernova 1987A (SN~1987A).
The explosions were modeled in 3D with the
   neutrino-hydrodynamics code {\bf\sc Prometheus-HOTB}, and light-curve
   calculations were performed in spherical symmetry with the
   radiation-hydrodynamics code {\bf\sc Crab}, starting at a stage of
   nearly homologous expansion.
Our results confirm the basic findings of the previous work:
   3D neutrino-driven explosions with SN~1987A-like energies
   synthesize an amount of $^{56}$Ni that is consistent with the radioactive tail of
   the light curve.
Moreover, the models mix hydrogen inward to minimum velocities below
   400\,km\,s$^{-1}$ as required by spectral observations and a 3D analysis
   of molecular hydrogen in SN~1987A.
Hydrodynamic simulations with the new progenitor models, which possess smaller
   radii than the older ones, show much better agreement between calculated
   and observed light curves in the initial luminosity peak and during the
   first 20 days.
A set of explosions with similar energies demonstrated that a high growth
   factor of Rayleigh-Taylor instabilities at the (C+O)/He composition
   interface combined with a weak interaction of fast Rayleigh-Taylor plumes,
   where the reverse shock occurs below the He/H interface, provides
   a sufficient condition for efficient outward mixing of $^{56}$Ni into
   the hydrogen envelope.
This condition is realized to the required extent only in one of the older
   stellar models, which yielded a maximum velocity of around 3000\,km\,s$^{-1}$
   for the bulk of ejected $^{56}$Ni, but failed to reproduce the helium-core
   mass of 6\,$M_{\sun}$ inferred from the absolute luminosity of the
   presupernova star.
We conclude that none of the single-star progenitor models proposed for
   SN~1987A to date satisfies all constraints set by observations.
}
%
\titlerunning{Mixing constraints on the SN~1987A progenitor}
\authorrunning{V. P. Utrobin et al.}
\maketitle
%
\section{Introduction}
\label{sec:intro}
%
The explosion of the blue supergiant (BSG) Sanduleak $-69^{\circ}202$ in
   the Large Magellanic Cloud (LMC) as the peculiar type II plateau supernova
   (SN) 1987A stimulated not only activity in its observation from radio
   wavelengths to gamma rays, but also the further development in the theory
   of the evolution of massive stars, the simulation of explosion mechanisms,
   and the modeling of light curves and spectra.
This well-observed object displayed a number of intriguing observational
   features \citep{ABKW_89}, two of which are interesting in the context of
   this paper.
First, the progenitor of SN~1987A was a compact star, but not
   a red supergiant (RSG).
Second, SN~1987A exhibited clear observational evidence for macroscopic
   mixing that must have occurred during the explosion.
   
Over the past three decades, the relative compactness of the BSG progenitor
   Sanduleak $-69^{\circ}202$ of SN~1987A has been a puzzling problem for
   the theory of the evolution of massive stars, and no consensus
   has been reached on its origin so far.
A single-star scenario to fit the observed properties of the progenitor
   with evolutionary calculations requires either a metal-deficient
   composition similar to the LMC \citep{Arn_87, HHTW_87}, a modification of
   convective mixing through rotation-induced meridional circulation in
   the star during its evolution \citep{WHT_88}, a restricted semiconvective
   diffusion \citep{WPE_88, Lan_91}, or both mass loss and convective mixing
   \citep{SNK_88}.
Binary stellar evolution invoking a strong accretion of matter
   from the companion star \citep{PJ_89} or a merger of the companion star
   with the primary RSG \citep{HM_89, PJR_90, PMI_07, MH_17} can also explain
   the properties of the SN~1987A progenitor.
\citet{Pod_92} showed, however, that only binary scenario models are able
   to fit all available observational and theoretical constraints.

Observational evidence for mixing of radioactive $^{56}$Ni and hydrogen in
   the ejected envelope exists not only at early times, but also at late times
   \citep[see][for details]{UWJM_15}.
During $20-100$ days after the explosion, the H$\alpha$ profile exhibited
   a striking fine structure called ``Bochum event'' \citep{HD_87, PH_89}.
At day 410, a unique high-velocity feature with a radial velocity of about
   $+3900$\,km\,s$^{-1}$ was found in the infrared emission lines of
   [\ion{Fe}{ii}] and was interpreted as a fast-moving iron clump \citep{HCE_90}.
\citet{UCA_95} analyzed the H$\alpha$ profile at the Bochum event phase and
   identified this high-velocity feature with a fast $^{56}$Ni clump that
   was moving at an absolute velocity of 4700\,km\,s$^{-1}$ and had a mass
   of $\sim$10$^{-3}\,M_{\sun}$.
The bulk of radioactive $^{56}$Ni had a maximum velocity of
   $\sim$3000\,km\,s$^{-1}$, as inferred from the infrared
   emission lines of [\ion{Ni}{ii}] and [\ion{Fe}{ii}] at day 640
   \citep{CHELH_94}.
\citet{Chu_91} studied the profiles of hydrogen emission lines at day 350
   and found that the slowest-moving hydrogen-rich matter had
   a velocity of $\sim$600\,km\,s$^{-1}$.
In turn, \citet{KF_98} analyzed the H$\alpha$ profile taken at day 804 and
   argued that hydrogen was mixed down into the innermost ejecta to velocities
   of $\le 700$\,km\,s$^{-1}$.
\citet{MJS_12} revisited the line profiles during the nebular phase and showed
   that they are peaked in shape, suggesting that mixing of the
   elements including hydrogen must have occurred in the ejecta down to zero
   velocity.
Reconstructing a 3D view of molecular hydrogen in SN~1987A, \citet{LSF_19}
   found that the lowest observed velocities of molecular hydrogen are in
   the range of 400 to 800\,km\,s$^{-1}$.
Thus, the presence of hydrogen in the innermost layers of the ejecta is
   confirmed by observational data of SN~1987A.

These comprehensive observational data unambiguously demonstrate that
   the envelope of the pre-SN was substantially mixed during the explosion.
In the framework of the standard paradigm of neutrino-powered explosions in
   2D geometry, \citet{KPSJM_06} demonstrated that massive
   $^{56}$Ni-dominated clumps can be mixed deep into the helium shell and
   even into the hydrogen layer of the disrupted star, with terminal velocities
   of up to $\sim$3000\,km\,s$^{-1}$.
In turn, hydrogen can be mixed at the He/H composition interface downward
   in velocity space to about 500\,km\,s$^{-1}$.
Later 3D hydrodynamic simulations of neutrino-driven
   explosions for SN~1987A \citep{HJM_10, WJM_10, MJW_12, WJM_13,
   WMJ_15} confirmed the basic results of the 2D simulations concerning
   the mixing of radioactive $^{56}$Ni and hydrogen.
\citet{KPSJM_06} and \citet{WMJ_15} found that the amount of outward $^{56}$Ni
   mixing and inward hydrogen mixing is sensitive to the structure of
   the helium core and the He/H composition interface.

Using this sensitivity, \citet{UWJM_15} studied the dependence of explosion
   properties on the structure of four BSG progenitors for the first time
   in the framework of the neutrino-driven explosion mechanism.
The authors compared their light curves with observations of SN~1987A.
The considered pre-SN models, 3D explosion simulations, and light-curve  
   calculations explain the basic observational features of SN~1987A.
However, only one hydrodynamic model reproduces the width of the broad maximum
   of the observed light curve promisingly well, and no model matches the
   observational features connected to the pre-SN structure of the outer
   stellar layers.
All progenitor models have radii that are too large to reproduce the observed
   narrow initial luminosity peak, and the structure of their outer layers
   does not match the observed light curve during the first 30--40 days.

We therefore explore three
   new BSG models for Sanduleak $-69^{\circ}202$ that have smaller radii
   than the old models, some of which were discussed in \citet{SEWBJ_16}.
Applying the self-consistent approach developed in \citet{UWJM_15} to these
   three new BSG models, we find that the mixing that occurs during the explosion
   plays a crucial role in constraining the helium-core properties of the
   progenitor.

We begin in Sect.~\ref{sec:modmeth} with a brief description of the pre-SN
   models and the numerical approach.
In Sect.~\ref{sec:results} we present and analyze our results, and compare them
   with SN~1987A observations.
We conclude in Sect.~\ref{sec:conclsn} with a summary and discussion of our
   findings.
In Appendix~\ref{sec:apndx} we discuss the role of 3D macroscopic mixing
   in light-curve modeling of ordinary and peculiar type IIP SNe.

\section{Model overview and numerical approach}
\label{sec:modmeth}
%
Our overview of pre-SN models includes seven (three new and four previously
   used) available models obtained in the scenario of single-star evolution.
The numerical approach we follow in this work is identical to that presented
   in \citet{UWJM_15}.
It consists of three steps: 3D simulations of the neutrino-driven explosion,
   mapping 3D simulation results to 1D, and hydrodynamic light-curve
   modeling.
We briefly review each step in the following subsections.

\subsection{Presupernova models}
\label{sec:modmeth-psn}
%
\begin{table}[t]
\caption[]{Presupernova models for blue supergiants.}
\label{tab:presnm}
\centering
\begin{tabular}{@{ } l @{ } c @{ } c @{ } c @{ } c @{ } c @{ } c @{ } c @{ } c @{ } c @{ }}
\hline\hline
\noalign{\smallskip}
 Model & $R_\mathrm{pSN}$ & $M_\mathrm{CO}^{\,\mathrm{core}}$
       & $M_\mathrm{He}^{\,\mathrm{core}}$ & $M_\mathrm{pSN}$
       & $M_\mathrm{ZAMS}$ & $Y_\mathrm{surf}$ & $Z_\mathrm{surf}$
       & Rot. & Ref. \\
\noalign{\smallskip}
       & $(R_{\sun})$ & \multicolumn{4}{c}{$(M_{\sun})$} &
       & $(10^{-2})$  &     &    \\
\noalign{\smallskip}
\hline
\noalign{\smallskip}
 W16   & 28.8 & 2.57 & 6.55 & 15.36 & 16    & 0.521 & 0.50 & Yes & 1 \\
 W18r  & 41.9 & 2.67 & 6.65 & 17.09 & 18    & 0.453 & 0.50 & Yes & -- \\
 W18x  & 30.4 & 2.12 & 5.13 & 17.56 & 18    & 0.281 & 0.60 & Yes & 1 \\
\noalign{\smallskip}
\hline
\noalign{\smallskip}
 B15   & 56.1 & 1.64 & 4.05 & 15.02 & 15.02 & 0.230 & 0.34 &  No & 2 \\
 W18   & 46.8 & 3.06 & 7.40 & 16.92 & 18.0  & 0.515 & 0.50 & Yes & 1 \\
 W20   & 64.2 & 2.33 & 5.79 & 19.38 & 20.10 & 0.256 & 0.56 &  No & 3 \\
 N20   & 47.9 & 3.75 & 5.98 & 16.27 & $\sim$20.0 & 0.435 & 0.50 &  No & 4 \\
\noalign{\smallskip}
\hline
\end{tabular}
\tablefoot{%
The columns list the name of the pre-SN model; its radius, $R_\mathrm{pSN}$;
   the CO-core mass, $M_\mathrm{CO}^{\,\mathrm{core}}$;
   the helium-core mass, $M_\mathrm{He}^{\,\mathrm{core}}$;
   the pre-SN mass, $M_\mathrm{pSN}$; the progenitor mass,
   $M_\mathrm{ZAMS}$; and the mass fraction of helium, $Y_\mathrm{surf}$,
   and heavy elements, $Z_\mathrm{surf}$,
   in the hydrogen-rich envelope at the stage of core collapse.
The last two columns note whether the progenitor was rotating or not, and
   the corresponding reference is listed.
The lower four initial models were previously used in \citet{UWJM_15}.
}
\tablebib{
(1) \citet{SEWBJ_16};
(2) \citet{WPE_88};
(3)~\citet{WHWL_97};
(4) \citet{SN_90}.
}
\end{table}
Most pre-SN models of our study are described in detail in \citet{SEWBJ_16}.
Here we only provide a brief overview of their basic properties
   (Table~\ref{tab:presnm}).
Model B15 (model W15 in the nomenclature of \citet{SEWBJ_16}) was evolved
   by \citet{WPE_88} from the main sequence to the precollapse stage.
This single 15\,$M_{\sun}$ star has a reduced metallicity.
When evolved with restricted semiconvection and without mass loss, it produced
   a BSG star.
Its luminosity, related to its helium-core mass\footnote%
     {We define the helium-core mass as the mass enclosed by the shell
     where the mass fraction of hydrogen $X$ drops below a value of $X=0.01$
     when moving inward from the surface of a star.
     The CO-core mass is determined in the same way, but for the mass fraction
     of helium $Y$ at a value of $Y=0.01$.}
   of 4.05\,$M_{\sun}$, is too low
   compared to the observational data of Sanduleak $-69^{\circ}202$,
   which require a single star to have a helium-core mass of
   about 6\,$M_{\sun}$ \citep{SNK_88, WPE_88}.

\begin{figure*}[t]
\centering
   \includegraphics[width=0.48\hsize, clip, trim=18 153 67 99]{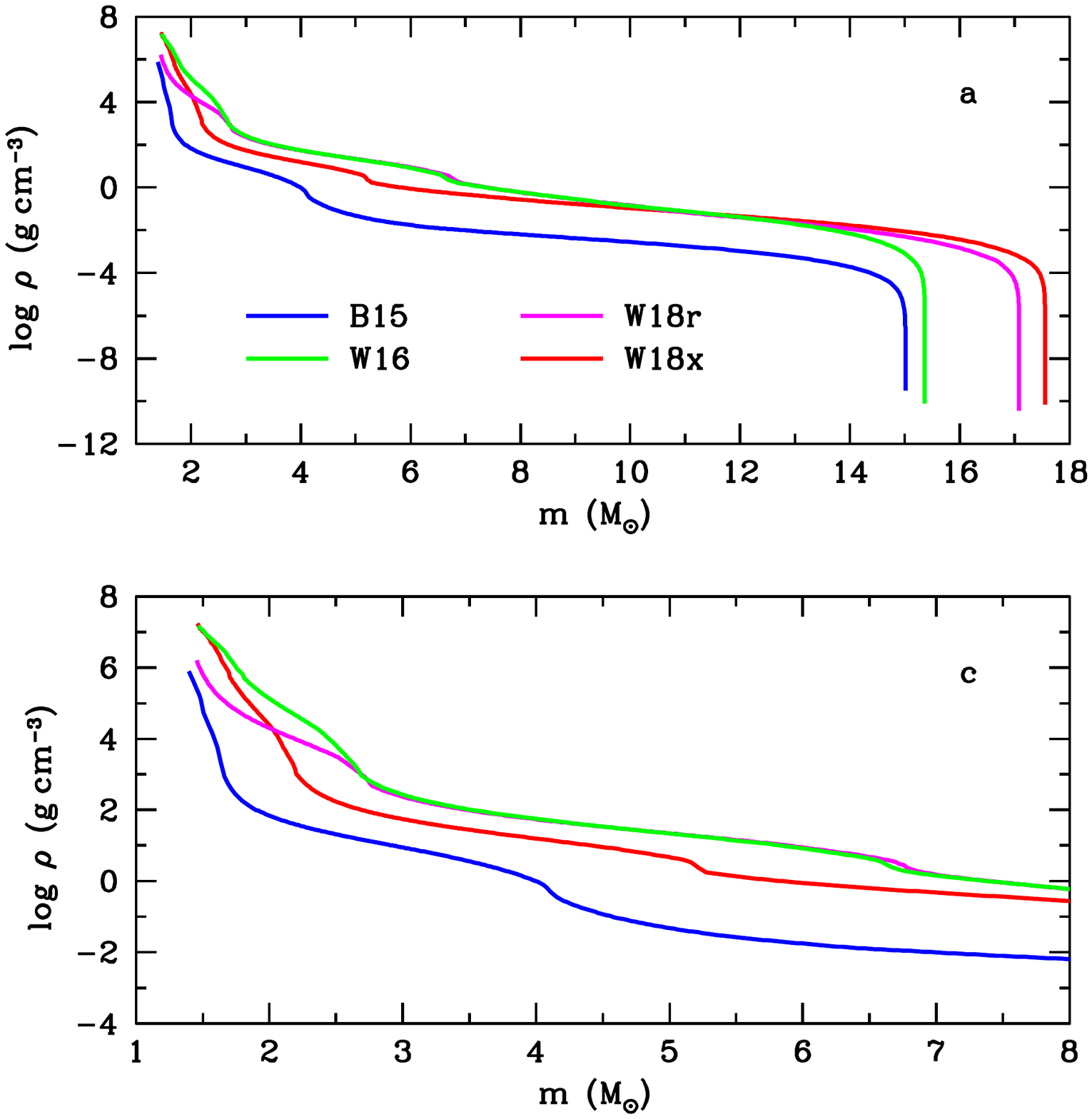}
   \hspace{0.5cm}
   \includegraphics[width=0.48\hsize, clip, trim=18 153 67 99]{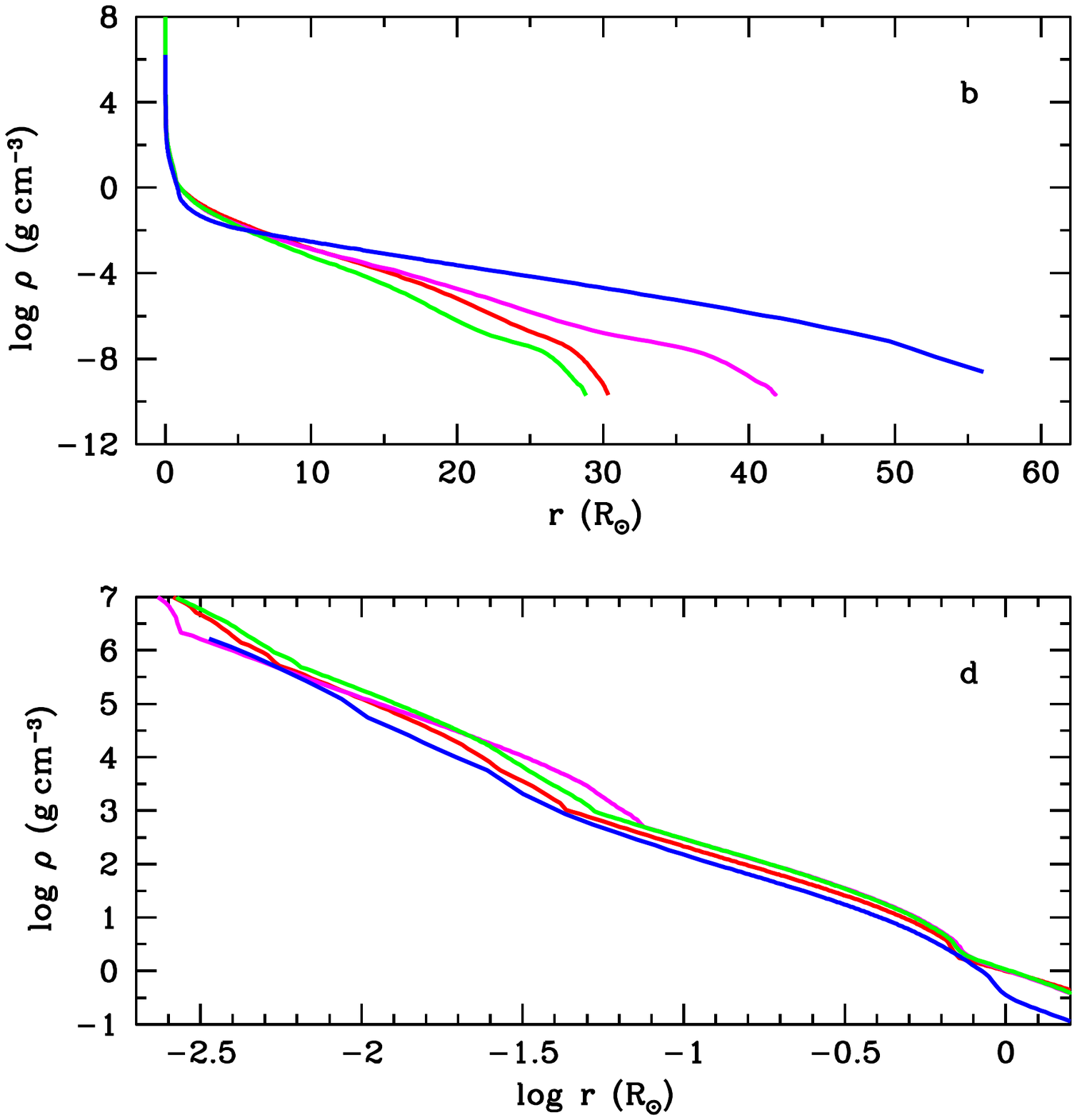}\\
   \caption{%
   Density distribution as a function of interior mass for the whole star
      (a) and the inner region of 8\,$M_{\sun}$ (c), and as a function of
      radius for the whole star (b) and the inner region of 1\,$R_{\sun}$ (d)
      in the pre-SN models B15, W16, W18r, and W18x.
   The central collapsing core that is integrated into the neutron star
      is omitted.
   }
   \label{fig:denmr}
\end{figure*}
We here study models W16, W18, and W18x of the BSG progenitor models presented
   in \citet{SEWBJ_16}.
Model W18 was evolved with reduced metallicity and restricted
   semiconvection, like model B15, and with a substantial amount of rotation
   and mass loss.
Rotation increases the helium-core mass to 7.4\,$M_{\sun}$ and also stirs
   more helium into the hydrogen envelope, enriching its surface mass fraction
   to 0.515.
A higher helium-core mass implies a higher luminosity, somewhat above the
   observed limit for Sanduleak $-69^{\circ}202$.
Two other models, which are similar to model W18, illustrate some sensitivities:
   model W16 is identical to model W18, except for its different mass and
   rotation rate; model W18x is slightly different in input physics and has
   less total angular momentum on the main sequence.
The angular momenta on the main sequence for models W16, W18, and W18x are
   $(2.7, 3.1, \mathrm{and}\ 2.3)\times10^{52}$\,erg\,s, respectively.

\begin{figure*}[t]
\centering
   \includegraphics[width=0.9\hsize, clip, trim=29 162 42 211]{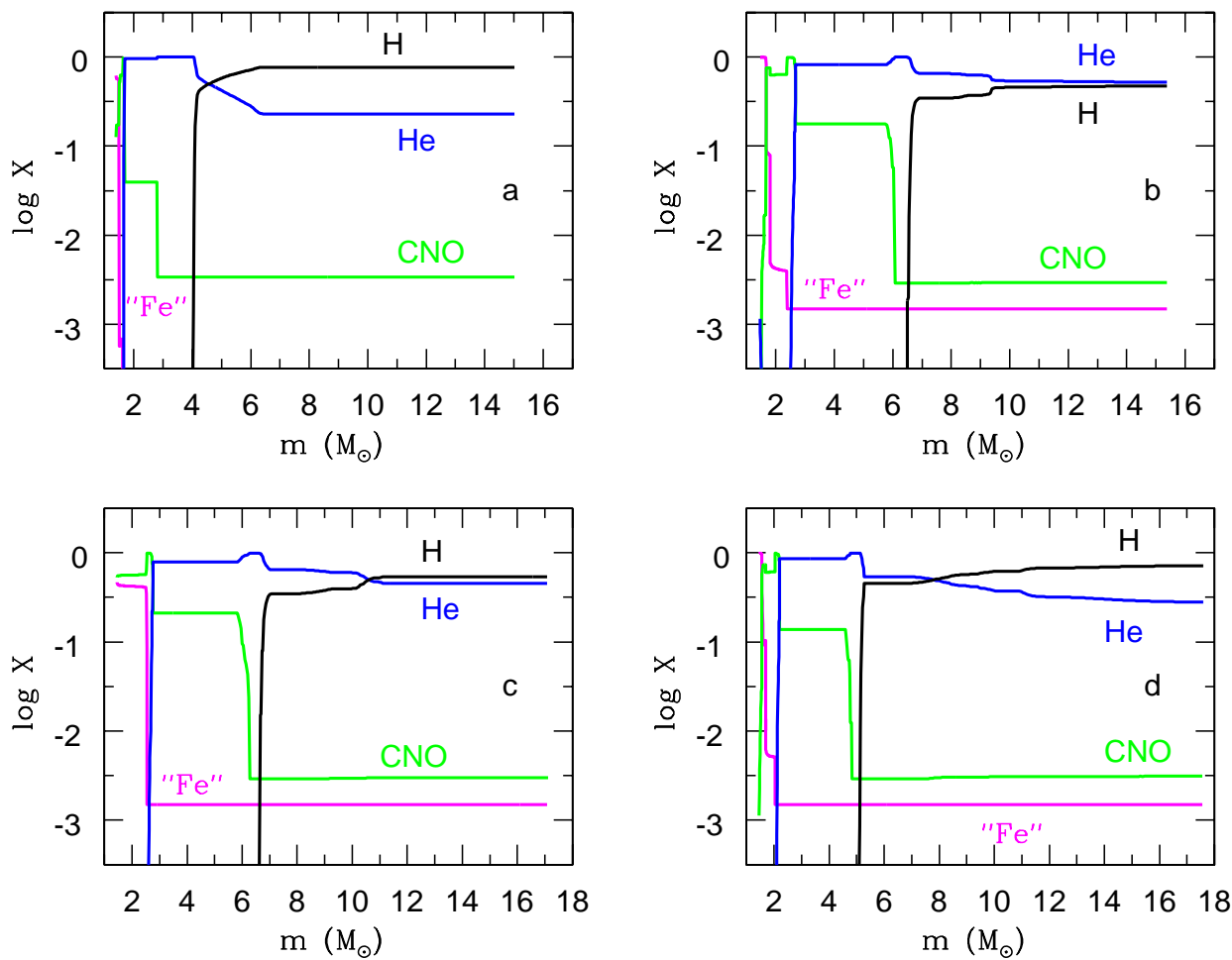}
   \caption{%
   Mass fractions of hydrogen (black), helium (blue), CNO-group elements
      (green), and iron-group elements (magenta) in the pre-SN
      models B15 (a), W16 (b), W18r (c), and W18x (d).
   }
   \label{fig:chcom}
\end{figure*}
Model W18r is an 18\,$M_{\sun}$ progenitor evolved with rotation and
   angular momentum transport, including magnetic torques \citep{Spr_02, HWS_05}.
The total initial angular momentum on the main sequence was
   $3.1\times10^{52}$\,erg\,s, which was reduced by transport and mass loss to
   $2.4\times10^{51}$\,erg\,s at death.
The initial mass was 18.0\,$M_{\sun}$, and the final mass was 17.09\,$M_{\sun}$,
   with a helium core of 6.65\,$M_{\sun}$ and a CO core of 2.67\,$M_{\sun}$.
The initial composition was taken to be representative of the subsolar
   metallicity of the LMC.
Hydrogen, helium, carbon, nitrogen, and oxygen had mass fractions of
   0.745, 0.250, $5.2\times10^{-4}$, $1.2\times10^{-4}$, and
   $2.5\times10^{-3}$, respectively.
We note that this initial composition was used in the other models W16, W18,
   W18x, and W20 as well.
These may not be the best modern values, but the total abundance of CNO-group
   elements is reduced by about a factor of three compared to solar.
This weakens the hydrogen-burning shell and helps to make a BSG star.

Model W20 was evolved by \citet{WHWL_97}, again including mass loss, and
   using the revised opacities of \citet{RI_92} and \citet{IR_96}, which
   were also applied in all other models except for B15 (where recalculation
   with the new opacities did not yield any significant differences). 
Its helium-core mass of 5.79\,$M_{\sun}$, the corresponding luminosity, and
   radius agree well with observations of Sanduleak $-69^{\circ}202$.
The hydrogen envelope is not enriched in helium, whose surface mass fraction
   is 0.256.

In all model progenitors described above, nuclear burning was followed using
   the usual 19 isotope ``approx'' network until oxygen depletion and
   the 128 isotope quasi-equilibrium network thereafter \citep{WZW_78}.
In addition to this standard reaction network, in model W18r, nucleosynthesis
   and gradual changes in the electron mole number were followed using
   an adaptive network \citep{RHHW_02} that contained about 1500 isotopes
   at the end.

Model N20 was constructed by \citet{SN_90} by combining the pre-SN helium core
   of 6\,$M_{\sun}$ evolved by \citet{NH_88} with the hydrogen envelope that was
   independently calculated by \citet{SNK_88}.
Both the helium core and the hydrogen envelope were defined to satisfy
   observational constraints from Sanduleak $-69^{\circ}202$.
This model is therefore somewhat artificial and is not an evolutionary pre-SN model
   that is evolved continuously from the main sequence onward.

It is noteworthy that the low mass of the CO core relative to the helium-core
   mass has proven important for obtaining a BSG pre-SN solution.
This was achieved in the {\bf\sc Kepler} code by limiting semiconvection to
   a relatively low efficiency and neglecting overshoot mixing.
The progenitor models B15, W16, W18, W18r, W18x, and W20 that were evolved in
   this approach to a BSG configuration have the mass ratio of the CO core
   to the helium core in the range of 0.392 to 0.414, while model N20
   originally constructed to obtain a RSG solution has this ratio of 0.627
   (Table~\ref{tab:presnm}).

We already analyzed the four models B15, W18, W20, and N20 in a previous paper
   \citep{UWJM_15} and now focus on the three new models W16, W18r, and W18x;
   we use model B15 as a reference case.
The new models have different helium-core masses close to the observationally
   required value of 6\,$M_{\sun}$ (Table~\ref{tab:presnm};
   Figs.~\ref{fig:denmr}a and c), which allows us to study the sensitivity
   of the amount of outward $^{56}$Ni mixing and inward hydrogen mixing to
   the structure of the helium core and the He/H composition interface.
In addition, these models also differ in their density distributions
   (Fig.~\ref{fig:denmr}) and chemical compositions (Fig.~\ref{fig:chcom}).
All new models have smaller surface radii than the older models
   (Table~\ref{tab:presnm}).

\subsection{Numerical methods}
\label{sec:modmeth-numrcs}
%
Our 3D neutrino-driven explosion simulations begin shortly after the stellar
   core has collapsed and a newly formed SN shock wave has propagated to a mass
   coordinate of approximately 1.25\,$M_{\sun}$ inside the iron core.
The evolution during core collapse and core bounce were computed in spherical
   symmetry and were provided to us by \citet{SEWBJ_16}.
After mapping the 1D post-bounce data onto a 3D grid, the 3D calculations
   were carried out with the explicit finite-volume Eulerian multifluid
   hydrodynamics code {\sc Prometheus} \citep{FAM_91, MFA_91a, MFA_91b}.
Details of the physics modules implemented into the {\sc Prometheus} code
   and our numerical setup have been described in \citet{WJM_13} for
   neutrino-driven explosion simulations and in \citet{WMJ_15} for
   simulations of the late-time evolution from approximately 1.3\,s after core
   bounce onward.
Nevertheless, we briefly summarize the input physics and numerical methods
   employed by our code as follows.

The {\sc Prometheus} code uses a dimensionally split version of the
   piecewise parabolic method \citep{CW_84} to solve the multidimensional
   hydrodynamic equations.
A fast and efficient Riemann solver for real gases \citep{CG_85} is used
   to compute numerical fluxes at cell boundaries. 
Inside grid cells, where a strong shock wave is present, we recompute the 
   inter-cell fluxes using an approximate Riemann solver \citep{Liou_96} 
   to prevent numerical artifacts known as the odd-even decoupling 
   \citep{Quirk_94}.
The yin-yang overlapping grid \citep{KS_04}, implemented into {\sc Prometheus}
   as in \citet{WHM_10}, is employed for efficient spatial discretization of
   the computational domain.
Newtonian self-gravity is taken into account by solving Poisson's equation
   in its integral form, using an expansion into spherical harmonics
   \citep{MS_95}.
In addition, a general relativistic correction of the monopole term of the
   gravitational potential is applied during the explosion simulations
   following \citet{SKJM_06} and \citet{AJS_07}.

To model the explosive nucleosynthesis approximately, a small $\alpha$-chain
   reaction network, similar to the network described in \citet{KPSJM_03},
   is solved.
In order to unambiguously determine the inward mixing of hydrogen, free
   protons, which are produced when neutrino-heated matter freezes out
   from nuclear statistical equilibrium, are distinguished from hydrogen
   originating from the hydrogen-rich stellar envelope by tagging them as
   different species in our multicomponent treatment of the stellar plasma.

The revival of the stalled SN shock and the explosion are triggered by imposing
   a suitable value of the neutrino luminosities at an inner radial grid boundary
   located at an enclosed mass of 1.1\,$M_{\sun}$, well inside the
   neutrinosphere.
Outside this boundary, which shrinks with time to mimic the contraction of
   the proto-neutron star, we model neutrino-matter interactions by solving
   the neutrino radiation transport equation in a ``ray-by-ray'' manner and
   in the gray approximation as described in \citet{SKJM_06}.
The explosion energy of the model is determined by the imposed isotropic
   neutrino luminosity, whose temporal evolution we prescribe as well, and
   by the accretion luminosity that results from the progenitor-dependent mass
   accretion rate and the gravitational potential of the contracting neutron
   star.
 
Our 3D calculations terminate at approximately one day after the explosion
   when the SN shock wave has swept through the entire progenitor star.
The further time evolution of the SN outburst beyond one day is modeled in
   one dimension.
To this end, we compute angle-averaged profiles of hydrodynamic quantities
   and chemical abundances of the 3D flow at chosen times and interpolate
   these profiles onto the Lagrangian (mass) grid used in the 1D simulations.
The resulting data are the initial conditions for the hydrodynamic
   modeling of the SN outburst.
Our 3D simulations of neutrino-driven explosions eliminate the need to initiate
   the explosion by a supersonic piston or a thermal and/or kinetic bomb.

The numerical modeling of the SN outbursts employs the implicit Lagrangian
   radiation hydrodynamics code {\sc Crab} \citep{Utr_04, Utr_07}.
It integrates the set of spherically symmetric hydrodynamic equations including
   self-gravity, and a radiation transfer equation in the gray approximation
   \citep[e.g.,][]{MM_84}.
The time-dependent radiative transfer equation, written in a comoving
   frame of reference to an accuracy of order $v/c$ ($v$ is the fluid velocity, and
   $c$ is the speed of light), is solved as a system of equations for the zeroth
   and first angular moments of the nonequilibrium radiation intensity.
This system of two moment equations is closed by calculating a variable
   Eddington factor directly, taking into account scattering of radiation
   in the SN ejecta.
In the inner optically thick layers of the ejecta, the diffusion of equilibrium
   radiation is treated in the approximation of radiative heat conduction.
The resulting set of equations is discretized spatially using the method of
   lines \citep[e.g.,][]{HNW_93, HW_96}.
The energy deposition of gamma rays with energies of about 1\,MeV from the
   decay chain $^{56}$Ni $\to ^{56}$Co $\to ^{56}$Fe is determined by solving
   the corresponding gamma-ray transport.
The equation of state, the mean opacities, and the thermal emission coefficient
   are calculated taking non-LTE and nonthermal effects into account.
In addition, the contribution of spectral lines to the opacity in a medium
   expanding with a velocity gradient is estimated using the generalized formula
   of \citet{CAK_75}.
We refer to \citet{UWJM_15} and references therein for details of the numerical
   setup.

\section{Results}
\label{sec:results}
%
\begin{table*}[t]
\caption[]{Basic properties of the 3D hydrodynamic models.}
\label{tab:3Dsim}
\centering
\begin{tabular}{@{ } l @{ } c @{ } c @{ } c @{ } c @{ } c @{ } c @{ } c @{ } c @{ } c @{ } c @{ } c @{ } c @{ } c @{ } c @{ } c @{ }}
\hline\hline
\noalign{\smallskip}
 Model & \phantom{m}$M_\mathrm{NS}$
       & \phantom{m}$M_\mathrm{env}$
       & \phantom{m}$E_\mathrm{exp}$
       & \phantom{m}$M_\mathrm{Ni}^{\,\mathrm{min}}$
       & \phantom{e}$M_\mathrm{Ni}^{\,\mathrm{max}}$
       & \phantom{m}$M_\mathrm{Ni}^{\,\mathrm{i}}$
       & \phantom{m}$M_\mathrm{Ni}^{\,\mathrm{f}}$
       & \phantom{m}$v_\mathrm{Ni}^{\,\mathrm{bulk}}$
       & \phantom{m}$\langle v \rangle_\mathrm{Ni}^\mathrm{tail}$
       & \phantom{m}$v_\mathrm{H}^{\,\mathrm{mix}}$
       & \phantom{m}$\delta M_\mathrm{H}^{\,\mathrm{mix}}$
       & \phantom{m}$\delta M_\mathrm{p}^{\,\mathrm{free}}$
       & \phantom{m}$\Delta M_\mathrm{H}^\mathrm{2000}$
       & \phantom{m}$t_\mathrm{map}$
       & \phantom{m}$t_\mathrm{SB}$\phantom{m} \\
\noalign{\smallskip}
       & \multicolumn{2}{c}{$(M_{\sun})$}
       & (B)
       & \multicolumn{4}{c}{$(10^{-2}\,M_{\sun})$}
       & \multicolumn{3}{c}{($10^3$\,km\,s$^{-1}$)}
       & \multicolumn{2}{c}{$(10^{-2}\,M_{\sun})$}
       & $(M_{\sun})$
       & \multicolumn{2}{c}{($10^{3}$\,s)} \\
\noalign{\smallskip}
\hline
\noalign{\smallskip}
W16-1   & 2.08 & 13.25 & 0.88 & 2.47 &  5.07 & 5.07 & 3.47 & 1.43 & 1.52 & 1.27 &  9.74 & 1.32 & 0.694 & 89.28 & 3.39 \\
W16-2   & 1.90 & 13.45 & 1.16 & 3.03 &  8.00 & 8.00 & 6.92 & 2.37 & 2.50 & 1.48 & 12.52 & 2.32 & 0.322 & 88.82 & 3.03 \\
W16-3   & 1.66 & 13.69 & 1.48 & 4.06 & 12.43 & 8.67 & 7.85 & 2.71 & 2.95 & 1.85 &  4.42 & 3.32 & 0.060 & 88.49 & 2.66 \\
W16-4   & 1.58 & 13.78 & 1.81 & 4.70 & 15.46 & 8.47 & 7.97 & 2.84 & 3.02 & 2.01 &  6.91 & 3.87 & 0.037 & 88.26 & 2.44 \\
W18r-1  & 1.43 & 15.65 & 1.05 & 1.77 &  7.01 & 7.01 & 6.33 & 1.00 & 1.04 & 1.42 &  1.62 & 2.54 & 0.868 & 90.47 & 4.72 \\
W18r-2  & 1.35 & 15.73 & 1.31 & 2.46 &  9.35 & 7.90 & 7.57 & 1.15 & 1.20 & 1.61 &  1.61 & 3.07 & 0.462 & 89.77 & 4.30 \\
W18r-3  & 1.31 & 15.77 & 1.59 & 2.93 & 11.35 & 7.88 & 7.69 & 1.43 & 1.52 & 1.78 &  1.84 & 3.31 & 0.146 & 89.08 & 3.94 \\
W18r-4  & 1.27 & 15.80 & 1.91 & 3.12 & 13.18 & 7.78 & 7.76 & 1.68 & 1.75 & 1.98 &  2.59 & 3.41 & 0.039 & 89.26 & 3.62 \\
W18x-1  & 1.57 & 15.97 & 1.21 & 2.79 &  9.06 & 9.06 & 7.02 & 2.36 & 2.47 & 1.20 &  7.87 & 2.30 & 1.354 & 89.45 & 3.72 \\
W18x-2  & 1.52 & 16.03 & 1.45 & 3.89 & 12.21 & 8.43 & 7.55 & 2.46 & 2.85 & 1.36 &  5.40 & 2.78 & 0.847 & 89.13 & 3.43 \\
\noalign{\smallskip}
\hline
\noalign{\smallskip}
B15-2   & 1.25 & 14.20 & 1.40 & 3.07 &  9.28 & 7.29 & 7.25 & 3.37 & 3.50 & 0.27 & 15.32 & 1.92 & 0.922 & 61.22 & 6.20 \\
W18     & 1.47 & 15.45 & 1.36 & 2.91 & 11.00 & 8.38 & 7.62 & 1.38 & 1.47 & 1.76 &  1.69 & 3.11 & 0.168 & 55.20 & 4.26 \\
W20     & 1.49 & 17.87 & 1.45 & 3.23 & 11.18 & 7.72 & 7.48 & 1.39 & 1.49 & 1.31 &  4.21 & 3.35 & 1.312 & 61.24 & 6.66 \\
N20-P   & 1.46 & 14.69 & 1.67 & 3.97 & 11.71 & 7.65 & 7.58 & 1.61 & 1.79 & 0.19 & 22.66 & 3.61 & 0.353 & 56.86 & 5.40 \\
\noalign{\smallskip}
\hline
\end{tabular}
\tablefoot{%
The 3D models are based on the corresponding pre-SN models given in
   Table~\ref{tab:presnm}.
The lower four models have been analyzed in \citet{UWJM_15}.
$M_\mathrm{NS}$ is the baryonic mass of the neutron star at the end of the 3D
   simulations;
   $M_\mathrm{env}$ is the ejecta mass;
   $E_\mathrm{exp}$ is the explosion energy;
   $M_\mathrm{Ni}^{\,\mathrm{min}}$ is the mass of radioactive $^{56}$Ni produced
      directly by our $\alpha$-chain reaction network;
   $M_\mathrm{Ni}^{\,\mathrm{max}}$ is the aggregate mass of directly produced
      $^{56}$Ni and tracer nucleus; and
   $M^{i}_{\mathrm{Ni}}$ is the initial $^{56}$Ni mass at the onset of
      light-curve modeling. It is either set to be equal to
      $M_\mathrm{Ni}^{\,\mathrm{max}}$ (in models W16-1, W16-2, W18r-1,
      and W18x-1) or specified to fit the observed luminosity in the
      radioactive tail when $M_\mathrm{Ni}^{\,\mathrm{max}}$ is sufficiently
      high (in the other models);
   $M^{f}_{\mathrm{Ni}}$ is the $^{56}$Ni mass left after fallback in the ejecta
      at day 150;
   $v_\mathrm{Ni}^{\,\mathrm{bulk}}$ is the maximum velocity of the bulk mass of
      $^{56}$Ni;
   $\langle v \rangle_\mathrm{Ni}^\mathrm{tail}$ is the mean velocity of the
      fast-moving $^{56}$Ni tail;
   $v_\mathrm{H}^{\,\mathrm{mix}}$ is the minimum velocity of hydrogen mixed into
      the He shell, specified at the level where the mass fraction of hydrogen
      $X$ drops to value of $X=0.01$;
   $\delta M_\mathrm{H}^{\,\mathrm{mix}}$ is the mass of hydrogen mixed into
      the He shell;
   $\delta M_\mathrm{p}^{\,\mathrm{free}}$ is the mass of free protons left over
      in the neutrino-heated ejecta; and
   $\Delta M_\mathrm{H}^\mathrm{2000}$ is the mass of hydrogen confined to the
      inner layers that is ejected with velocities lower than 2000\,km\,s$^{-1}$.
$t_\mathrm{map}$ is the time at which the 3D simulation data are mapped onto
   the spherically symmetric grid.
$t_\mathrm{SB}$ is the epoch of shock breakout in the 1D simulations.
}
\end{table*}
We performed ten 3D neutrino-driven explosion simulations with the three new
   pre-SN models W16, W18r, and W18x (Table~\ref{tab:presnm}) as initial data.
In Table~\ref{tab:3Dsim} we list some basic properties of these 3D hydrodynamic
   models that we extracted at the end of the simulations.
For completeness, we also list the properties of the four old models B15-2,
   W18, W20, and N20-P that have been analyzed in \citet{UWJM_15}.
We define the explosion energy, $E_\mathrm{exp}$, as the sum of the total (i.e.,
   internal plus kinetic plus gravitational) energy of all grid cells
   at the mapping moment.
Throughout this paper, we employ the energy unit
   $1\,\mathrm{bethe}=1\,\mathrm{B}=10^{51}$\,erg.

\subsection{Production of $^{56}$Ni in neutrino-driven simulations}
\label{sec:results-niprod}
%
Our 3D supernova simulations are characterized by the explosion energy,
   the total amount of radioactive $^{56}$Ni, and the amount of macroscopic
   mixing of $^{56}$Ni and hydrogen-rich matter occurring during the SN
   explosion (Table~\ref{tab:3Dsim}).
To follow the explosive nucleosynthesis, we solved a small $\alpha$-chain
   reaction network and are therefore unable to determine the mass fraction
   of $^{56}$Ni in the tracer matter very accurately.
A possible overall production of radioactive $^{56}$Ni falls in between the
   minimum and maximum values: the mass of $^{56}$Ni produced directly by
   our $\alpha$-chain reaction network, $M_\mathrm{Ni}^{\,\mathrm{min}}$,
   and the aggregate mass of directly produced $^{56}$Ni and tracer nucleus,
   $M_\mathrm{Ni}^{\,\mathrm{max}}$ \citep[see][for details]{UWJM_15}).
Two sets of 3D models $\lbrace$W16-1, W16-2, W16-3, W16-4$\rbrace$ and
   $\lbrace$W18r-1, W18r-2, W18r-3, W18r-4$\rbrace$ show that the production
   of radioactive $^{56}$Ni is proportional to the explosion energy
   (Table~\ref{tab:3Dsim}); this confirms the results of \citet{UWJM_15}.
Moreover, the 3D models B15-2, W16-3, W18, W18r-2, W18x-2, W20, and N20-P
   comply with the finding that the $^{56}$Ni production is mostly dependent on the explosion
   energy (which is similar in all cases of this group of models), whereas
   the detailed properties of the pre-SN, that is, different helium-core masses
   ranging from 4 to 7.4\,$M_{\sun}$ (Table~\ref{tab:presnm}),
   different density structures (Fig.~\ref{fig:denmr};
   \citealt[Fig.~1]{UWJM_15}), and different chemical compositions
   (Fig.~\ref{fig:chcom}; \citealt[Fig.~2]{UWJM_15}), play a less important
   role.

\subsection{Mixing in neutrino-driven explosion simulations}
\label{sec:results-3Dexp}
%
\begin{figure*}[t]
\centering
   \includegraphics[width=0.9\hsize, clip, trim=38 151 48 97]{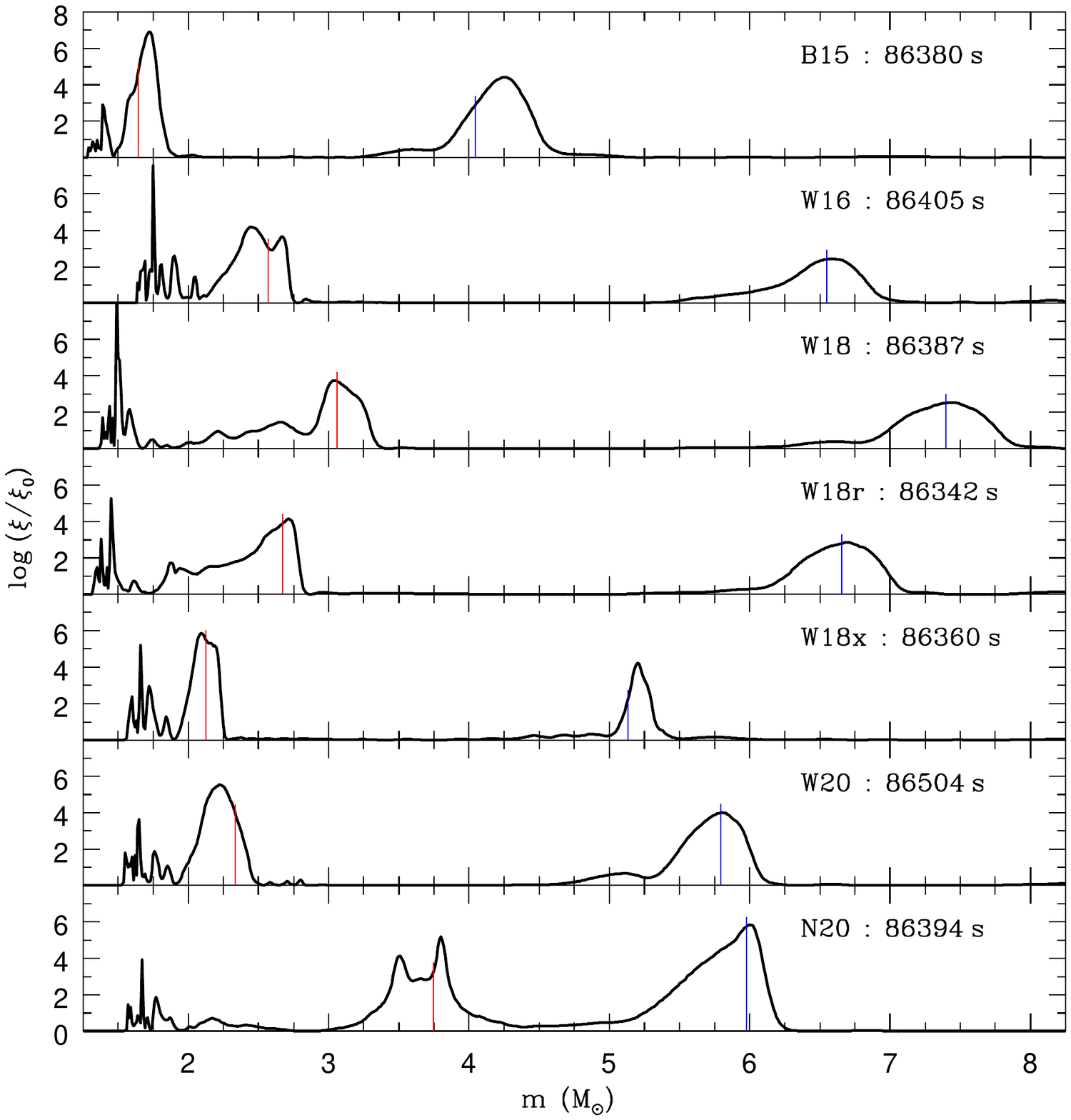}
   \caption{%
   Time-integrated Rayleigh-Taylor growth factors vs. enclosed mass for
      the 1D hydrodynamic models B15, W16, W18, W18r, W18x, W20, and N20
      at the given times.
   The vertical red and blue lines denote the mass coordinates of the
      (C+O)/He and He/H composition interfaces, respectively.
   These interfaces are Rayleigh-Taylor unstable after the passage of the
      SN shock.
   }
   \label{fig:rtgrowth}
\end{figure*}
In order to picture turbulent mixing in 3D simulations, we outline
   the development of neutrino-driven explosions after core bounce
   \citep[see, e.g.,][for details]{WMJ_15}.
The value of the explosion energy is determined by the isotropic neutrino
   luminosity at the inner grid boundary, its prescribed temporal evolution,
   and the accretion luminosity.
Neutrino heating causes buoyant bubbles to form in the convectively unstable
   layer that builds up within the neutrino-heating region between the gain
   radius and the stalled SN shock.
These high-entropy bubbles start to rise, grow, and merge.
As a result, the delayed, neutrino-driven explosion sets in, supported by
   convective overturn and large-scale aspherical shock oscillations
   caused by the standing accretion shock instability (SASI).

After the SN shock wave has been launched by neutrino heating, the further
   evolution of the explosion depends strongly on the density profile of the
   progenitor.
The shock decelerates when it encounters a density profile that falls off
   less steeply than $\rho \sim r^{-3}$, and it accelerates for density
   profiles that are steeper \citep{Sed_59}.
At the locations of the Si/O, (C+O)/He, and He/H composition interfaces
   (Fig.~\ref{fig:chcom}), the value of $\rho r^3$ varies nonmonotonically with
   radius such that the shock velocity increases when the shock approaches
   a composition interface and decreases after the shock has crossed
   the interface.
A deceleration of the shock causes a density inversion in the post-shock flow,
   which means that a dense shell forms.
Such shells at the locations of the composition interfaces are subject to
   Rayleigh-Taylor instabilities because they are characterized by density
   and pressure gradients of opposite signs \citep{Che_76}.

To compare the relative strength of the growth of Rayleigh-Taylor instabilities
   in different progenitor models, we performed an additional set of
   1D neutrino-driven explosion simulations for all progenitor stars,
   including those already presented in \citet{UWJM_15}.
Our 1D modeling approach is the same as in our 3D simulations.
This set of 1D models develops approximately the same explosion energy of
   1.4\,B.
To qualitatively analyze these instabilities in our 3D simulations, we
   computed the linear Rayleigh-Taylor growth rate in these 1D models for
   an incompressible fluid given by \citep{Ban_84, BT_90, MFA_91b}
\begin{equation}
   \sigma_{\mathrm{RT}} = \sqrt{-{\partial P \over \partial r}
      {\partial \rho \over \partial r}} \; .
\label{eq:sigmaRT}
\end{equation}
The growth factor for the time-dependent perturbation $\xi$ increases with 
   time $t$ as \citep{HTM_86}
\begin{equation}
   {\xi \over \xi_0}(t) = \exp\left(\int_{0}^t \sigma_{\mathrm{RT}}(\tau)\,d\tau
      \right) \; ,
\label{eq:rel_xi}
\end{equation}
   where $\xi_0$ is the amplitude of an initial perturbation at fixed
   Lagrangian mass coordinate.

Figure~\ref{fig:rtgrowth} displays the time-integrated Rayleigh-Taylor growth
   factors as functions of enclosed mass in our 1D models for each of the
   progenitors (Table~\ref{tab:presnm}) at times long after the shock breakout.
The growth factors in the Rayleigh-Taylor unstable layer near the (C+O)/He
   composition interface greatly vary between the different progenitors, with
   model B15 showing the largest growth factor.
On the other hand, the Rayleigh-Taylor growth factors at the He/H interface
   show less variation and differ only by approximately one order of magnitude
   between progenitors, except for model N20, which exhibits a particularly
   large growth factor. 
It is interesting to note here again that the helium core and the hydrogen
   envelope of the progenitor model N20 were evolved separately in two stellar
   evolution calculations by \citet{NH_88} and \citet{SNK_88}.
Thus, the large growth factor at the He/H interface observed in this model
   might not be physical.
It should also be emphasized that the results of the linear perturbation theory
   presented in Fig.~\ref{fig:rtgrowth} can only provide qualitative information
   on the relative strength of the expected growth of Rayleigh-Taylor
   instabilities in different layers of the progenitor star.
In realistic 3D simulations, the instabilities will quickly enter the nonlinear
   regime.
Nevertheless, the results from the linear analysis prove to be useful for
   a qualitative understanding of differences in the extent of mixing of
   $^{56}$Ni in different progenitors (see Sect.~\ref{sec:results-1Dpran}).

\begin{figure*}
\centering
   \includegraphics[width=0.245\hsize, clip, trim=60 60 60 60]{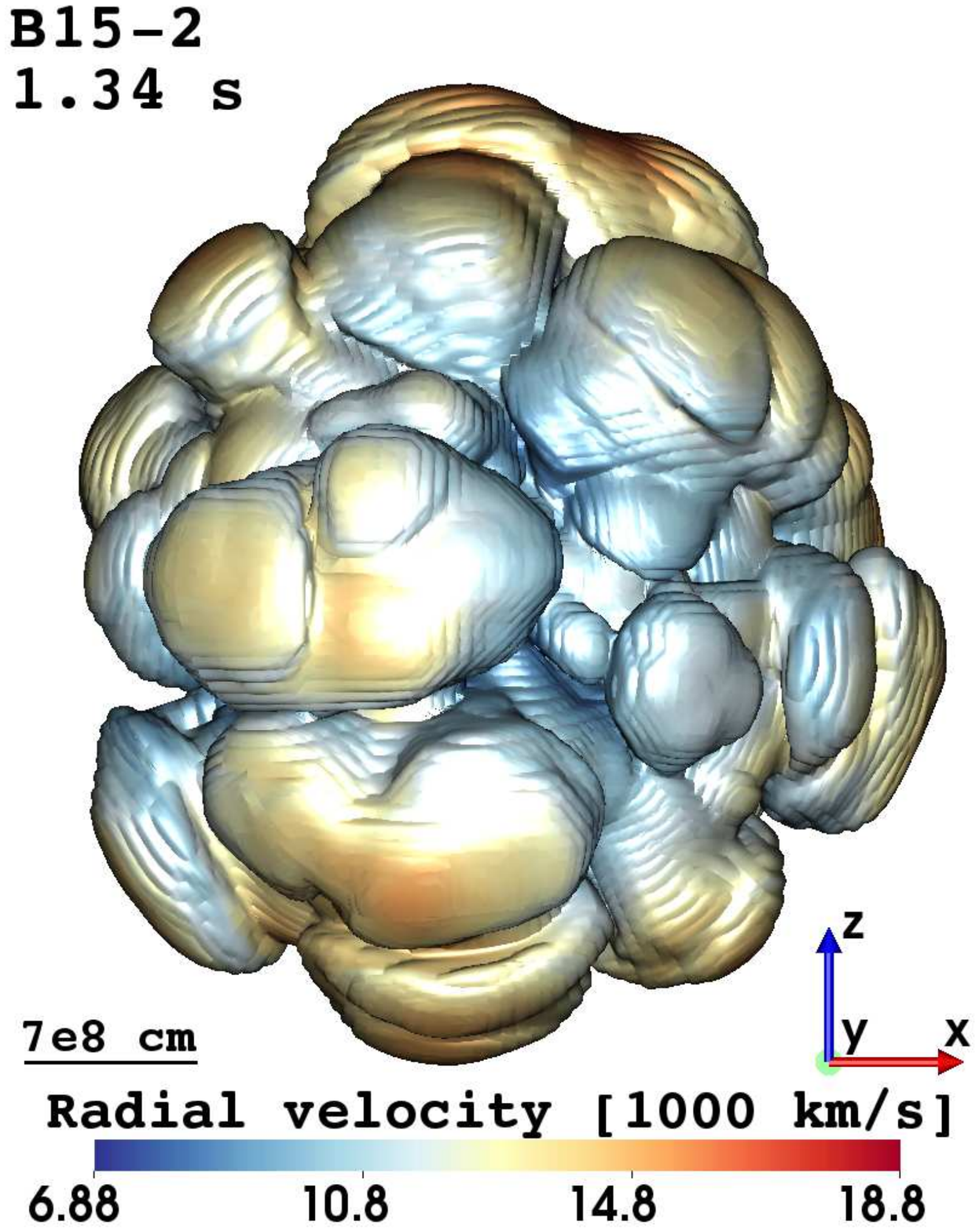}
   \includegraphics[width=0.245\hsize, clip, trim=60 60 60 60]{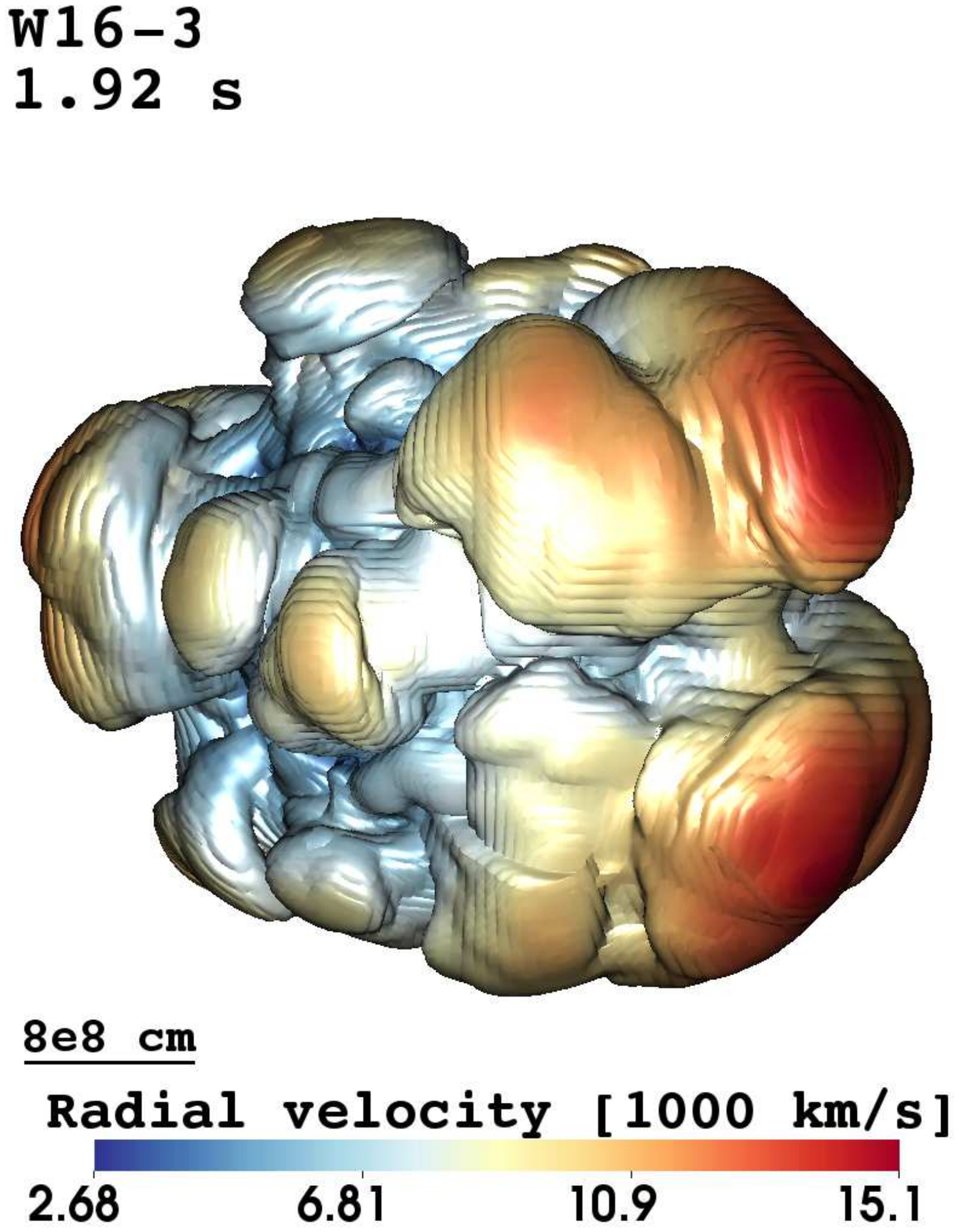}
   \includegraphics[width=0.245\hsize, clip, trim=60 60 60 60]{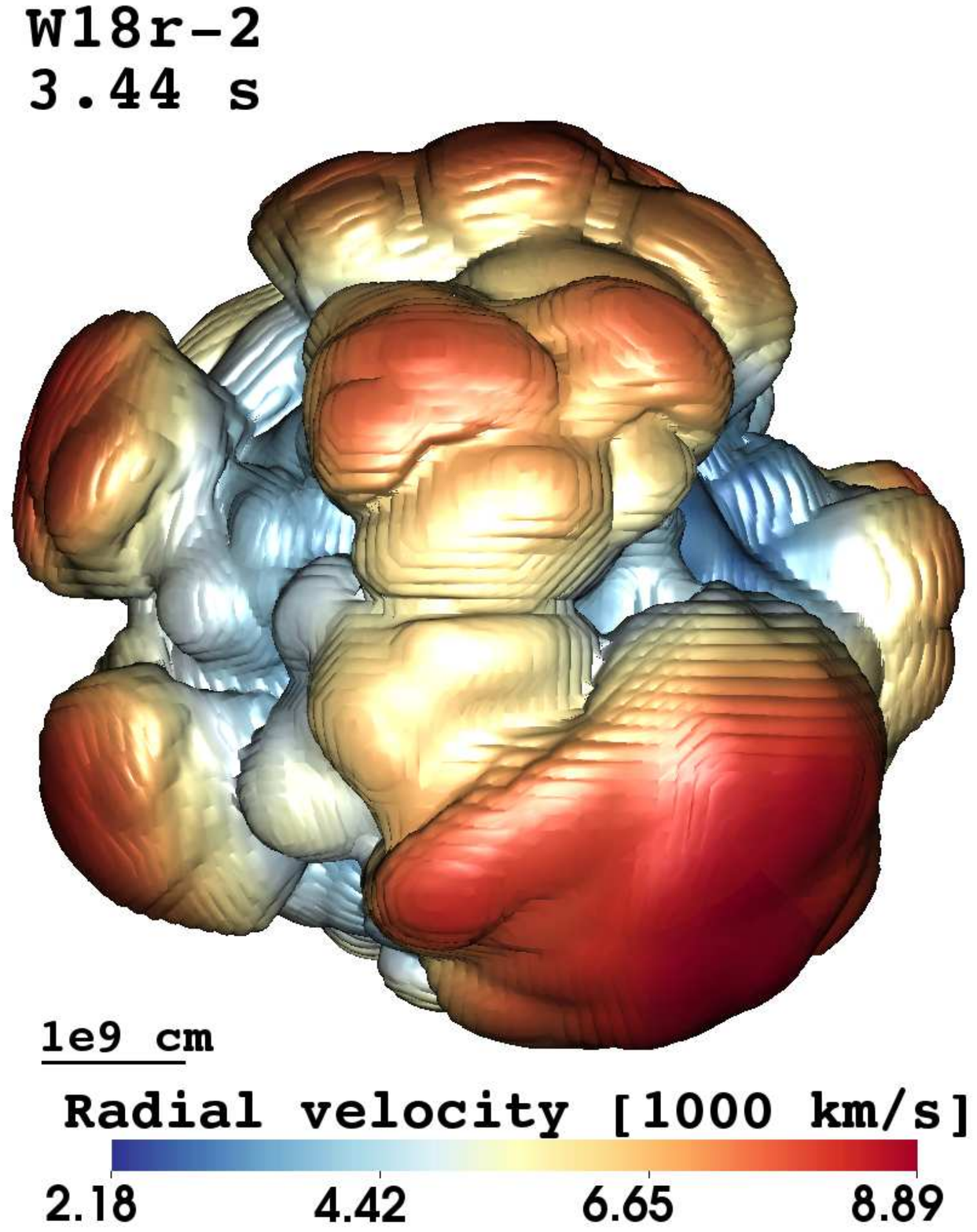}
   \includegraphics[width=0.245\hsize, clip, trim=60 60 60 60]{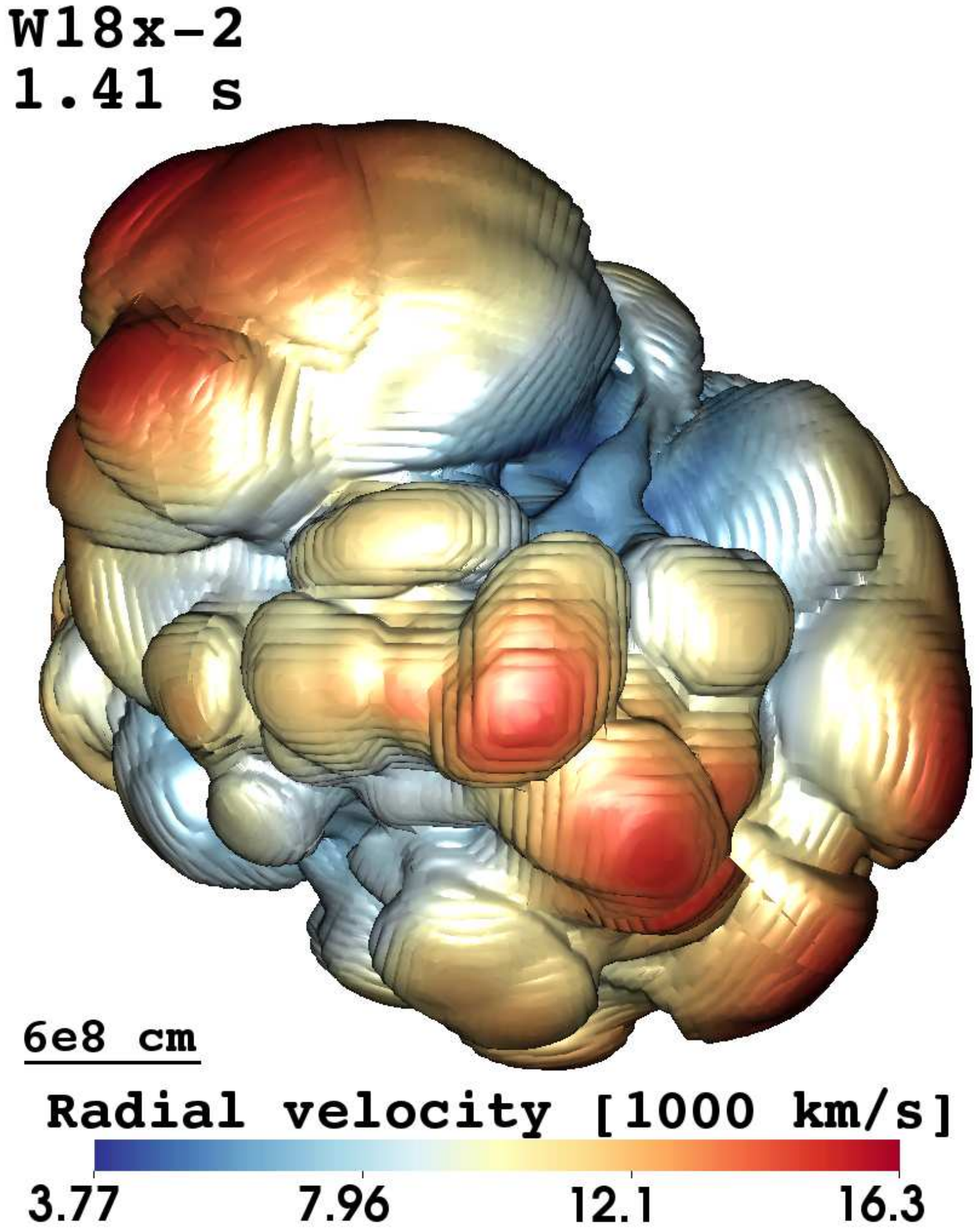}
   \hspace{0.0cm}
   \includegraphics[width=0.245\hsize, clip, trim=60 60 60 60]{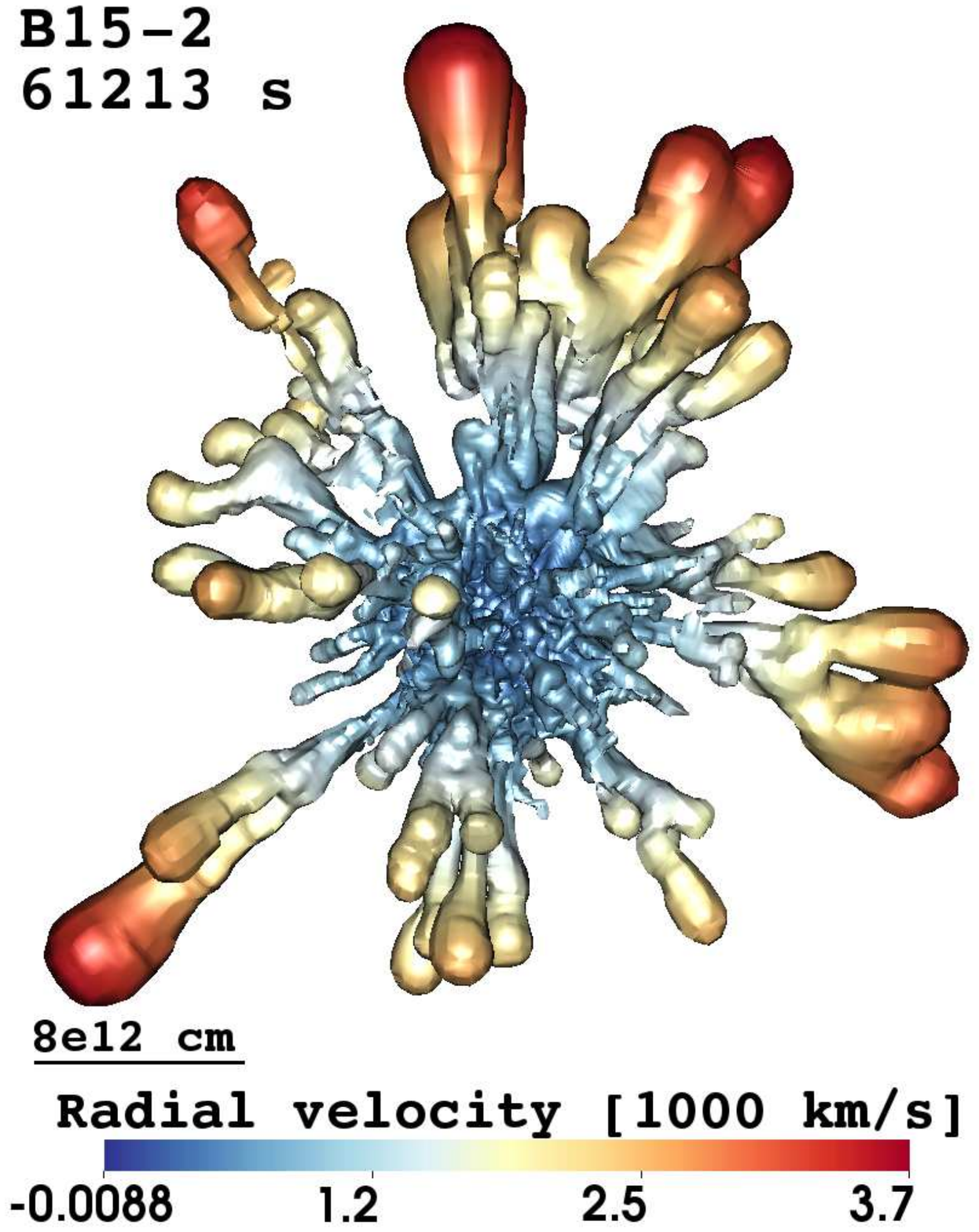}
   \includegraphics[width=0.245\hsize, clip, trim=60 60 60 60]{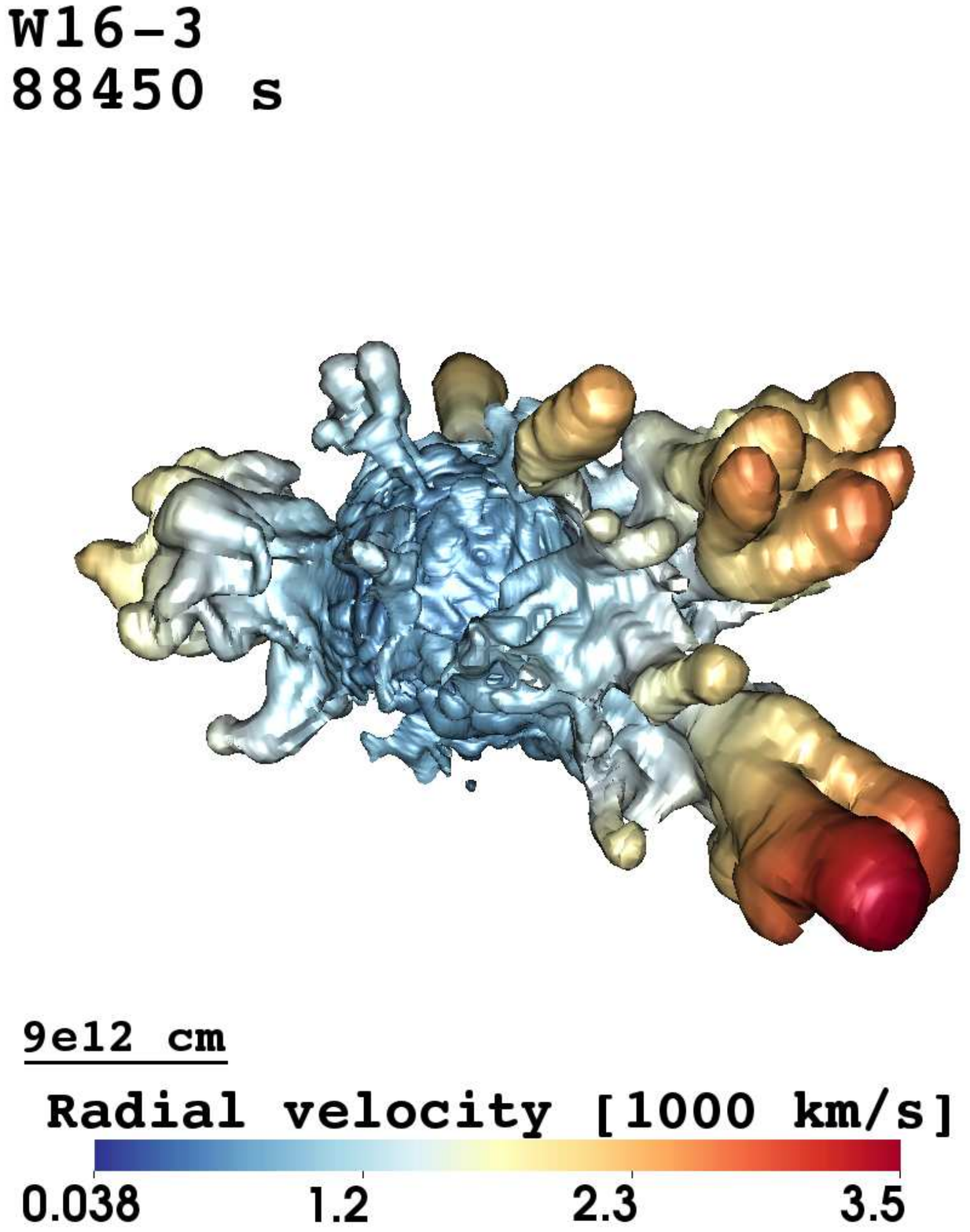}
   \includegraphics[width=0.245\hsize, clip, trim=60 60 60 60]{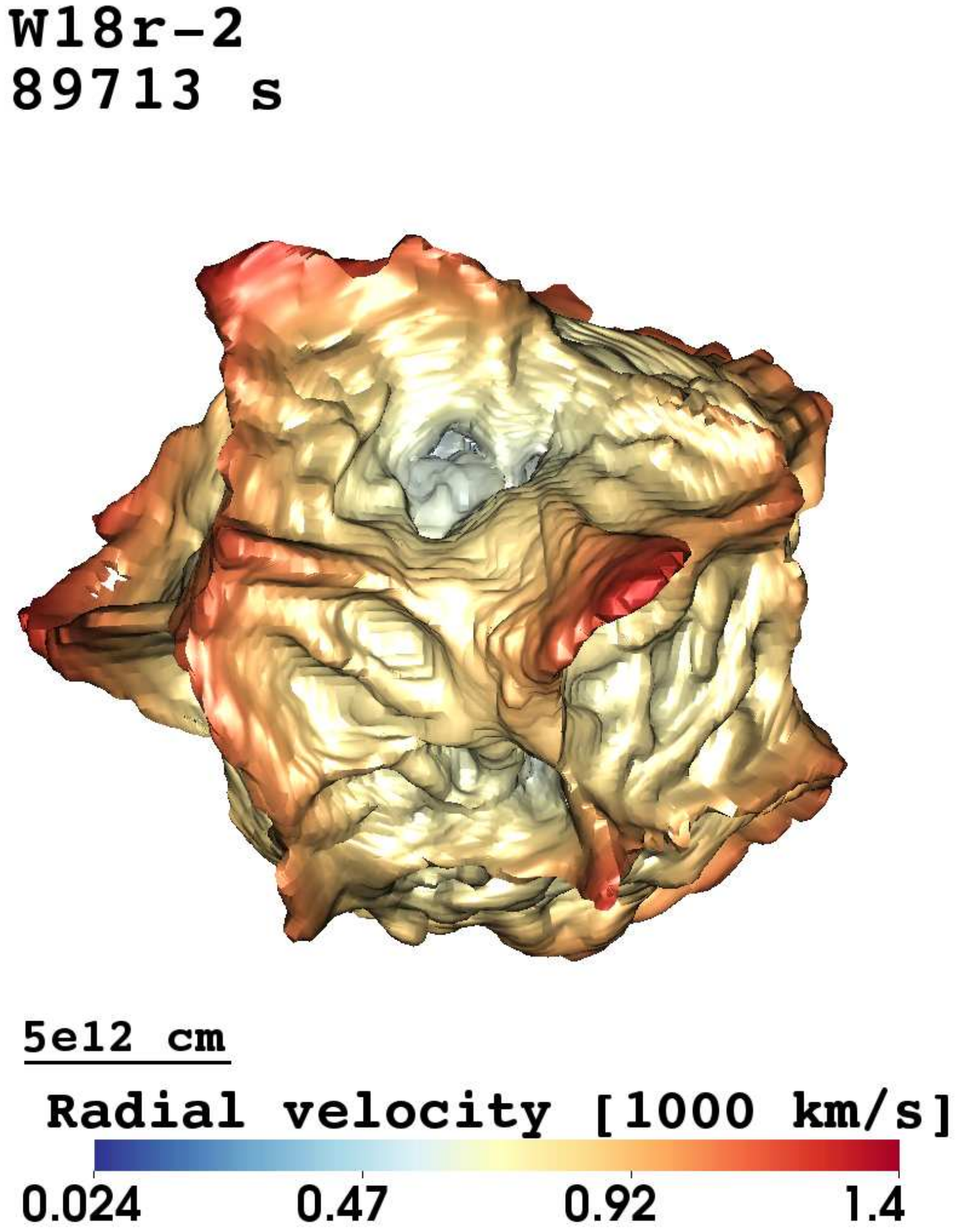}
   \includegraphics[width=0.245\hsize, clip, trim=60 60 60 60]{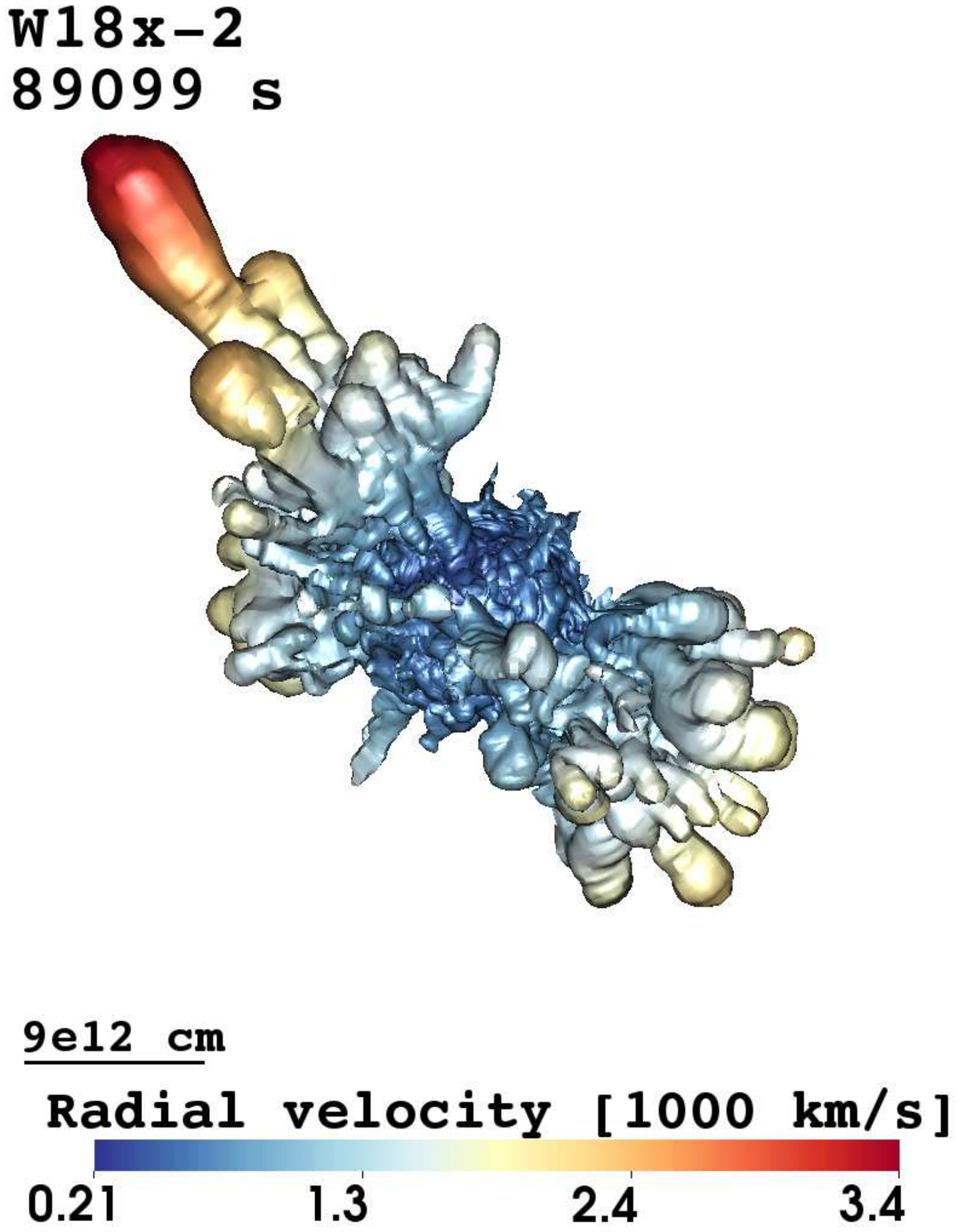}
   \caption{%
   Morphology of radioactive $^{56}$Ni-rich matter produced by explosive
      burning in shock-heated ejecta.
   The snapshots display isosurfaces where the mass fraction of $^{56}$Ni plus
      the neutron-rich tracer nucleus equals $3\%$.
   The isosurfaces are shown for four 3D models at two different epochs:
      shortly before the SN shock crosses the (C+O)/He composition interface
      in the progenitor star (upper row), and long after the shock breakout
      (lower row).
   In the top left corner of each panel, we give the name of the model and
      the post-bounce time of the snapshot.
   The colors give the radial velocity on the isosurface, and the color-coding
      is defined at the bottom of each panel.
   The size of the displayed volume and of the clumps can be estimated from
      the yardsticks given in the lower left corner of each panel.
   There is a striking difference between model B15-2 and the other three
      models in the final morphology of the $^{56}$Ni-rich ejecta, which
      arises from their specific progenitor structures and the influence of
      the latter on the unsteady SN shock propagation.
   }
   \label{fig:3D_models}
\end{figure*}
\begin{figure*}[t]
\centering
   \includegraphics[width=0.9\hsize, clip, trim=29 162 42 211]{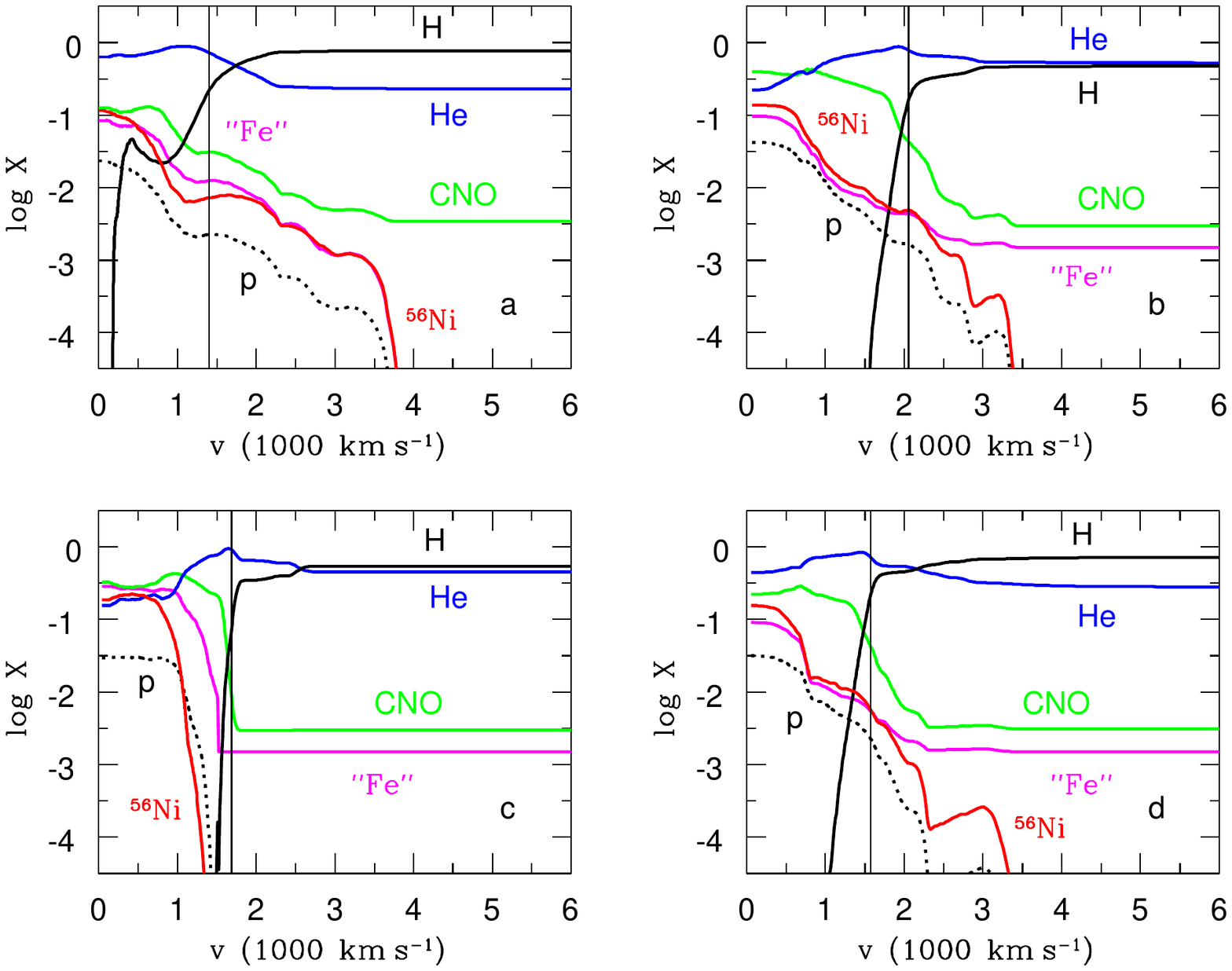}
   \caption{%
   Mass fractions of hydrogen (solid black), free protons (dotted black),
      helium (blue), CNO-group elements (green), iron-group elements except
      for $^{56}$Ni (magenta), and radioactive $^{56}$Ni (red) as functions
      of velocity at day 50 in models B15-2 (a), W16-3 (b), W18r-2 (c),
      and W18x-2 (d).
   Thin vertical lines indicate ejecta velocities at the location of
      the outer edge of the pre-SN helium core.
   }
   \label{fig:mfvel}
\end{figure*}

The SN shock wave first crosses the Si/O interface and then reaches the
   (C+O)/He interface, at which time the maximum speed of the bulk mass of
   $^{56}$Ni-rich matter, $v_\mathrm{Ni}^{\mathrm{CO}}$, 
   is spread over a wide range from 7850 to 19\,850\,km\,s$^{-1}$
   (Table~\ref{tab:nimixing}; Fig.~\ref{fig:3D_models}, upper row),
   depending on the progenitor, for models with comparable explosion
   energies (Table~\ref{tab:3Dsim}).
When the shock wave crosses the (C+O)/He interface and then enters the helium
   layer, a reverse shock forms.
After the main shock has passed the He/H interface, the evolution resembles
   the situation after crossing of the (C+O)/He interface, and another reverse
   shock forms below the He/H interface.
By this time, the fastest Rayleigh-Taylor plumes (a few in the case of
   model B15-2, and one plume in the case of models W16-3 and W18x-2) remain close 
   to the main shock and avoid interaction with the strong reverse shock.
Therefore, they penetrate the hydrogen envelope with high velocities,
   in stark contrast to the situation in the other models W18, W18r-2,
   W20, and N20-P.
As a result, the maximum velocities of iron-group nuclei and other elements
   decrease only slightly at late times, and the fastest $^{56}$Ni-rich clumps
   move with velocities of $\approx$3200 to $\approx$3700\,km\,s$^{-1}$
   in models W18x-2, W16-3, and B15-2
   (Fig.~\ref{fig:3D_models}, lower row; Fig.~\ref{fig:mfvel};
   \citealt[Fig.~3, lower row; Fig.~10]{UWJM_15}).
In contrast, the other mushrooms penetrate the reverse shock and move
   supersonically relative to the ambient medium, dissipating a large portion
   of their kinetic energy in bow shocks and strong acoustic waves, and forming
   an almost spherical distribution with velocities between $\approx$1100 and
   $\approx$2100\,km\,s$^{-1}$ in models W18r-2, W18, W20, and N20-P
   (Fig.~\ref{fig:3D_models}, lower row; Fig.~\ref{fig:mfvel};
   \citealt[Fig.~3, lower row; Fig.~10]{UWJM_15}).
This striking difference in the morphology of the $^{56}$Ni-rich ejecta was
   discussed extensively in previous works of \citet{KPSJM_03,KPSJM_06}.

The Rayleigh-Taylor instabilities and the global deformation of the main shock
   give rise to substantial inward mixing of hydrogen.
The aspherical shock deposits large amounts of vorticity into the layer at
   the He/H interface and thus induces mixing of hydrogen-rich matter down into
   the He shell.
To estimate the real extent of hydrogen mixing, we distinguish between
   hydrogen-rich matter (mass $\delta M_\mathrm{H}^{\,\mathrm{mix}}$) mixed
   from the hydrogen envelope into the He shell and free protons (mass
   $\delta M_\mathrm{p}^{\,\mathrm{free}}$) left uncombined in the
   neutrino-heated ejecta originating from the SN explosion
   (Table~\ref{tab:3Dsim}, Fig.~\ref{fig:mfvel}).

In Table~\ref{tab:3Dsim} we present some global properties of mixing induced
   by the 3D neutrino-driven explosions in the ejecta.
To characterize mixing of radioactive $^{56}$Ni in radial velocity space,
   we divided the $^{56}$Ni-rich ejecta into two components: a slow-moving
   bulk of $^{56}$Ni containing $99\%$ of the total $^{56}$Ni
   mass, and a fast-moving $^{56}$Ni tail containing the remaining $1\%$.
This division is motivated by the observational evidence for a fast $^{56}$Ni
   clump of $\sim$10$^{-3}\,M_{\sun}$ in SN~1987A \citep{UCA_95}.
We determined the maximum velocity of the bulk component,
   $v_\mathrm{Ni}^{\,\mathrm{bulk}}$, and the mean velocity of the tail
   component, $\langle v \rangle_\mathrm{Ni}^\mathrm{tail}$.
On the other hand, inward mixing of hydrogen-rich matter may be measured by
   the mass of hydrogen mixed into the He shell and its minimum velocity,
   $v_\mathrm{H}^{\,\mathrm{mix}}$.
We also evaluated the total mass of free protons produced in the
   neutrino-heated matter and left over in the SN ejecta.
Finally, it is instructive to determine the mass of hydrogen confined to
   the inner layers ejected with velocities lower than 2000\,km\,s$^{-1}$,
   $\Delta M_\mathrm{H}^\mathrm{2000}$, which was estimated by \citet{KF_98}.

Two sets of new models $\lbrace$W16-1, W16-2, W16-3, W16-4$\rbrace$, 
   $\lbrace$W18r-1, W18r-2, W18r-3, W18r-4$\rbrace$ (Table~\ref{tab:3Dsim}),
   and a set of old models B15-1, B15-2, B15-3 \citep{UWJM_15} demonstrate
   that the greater the explosion energy, the more intense
   the outward mixing of radioactive $^{56}$Ni in velocity space.
In turn, the mass of hydrogen mixed into the He shell in the two sets of
   new models does not correlate with the explosion energy, while its minimum
   velocity grows with the explosion energy (Table~\ref{tab:3Dsim}).
The finding regarding the mass of hydrogen mixed into the He shell has no
   simple explanation because the large-scale mixing is the result
   of a complex sequence of hydrodynamic instabilities at the composition
   interfaces \citep[cf.][]{WMJ_15}.

The mass of free protons left over in the neutrino-heated ejecta from
   the 3D explosion simulations increases with the explosion energy
   (Table~\ref{tab:3Dsim}).
Their distributions in velocity space are very similar to those of radioactive
   $^{56}$Ni (Fig.~\ref{fig:mfvel}) because the origins of the two
   nucleosynthesis components are similar.
The extent of free proton mixing may be measured by the maximum velocity of
   the bulk mass of $^{56}$Ni.
The smoothness of the resulting distribution of hydrogen matter is determined
   by overlapping the distributions of hydrogen mixed into the He shell and
   these free protons.
The degree of this overlapping correlates with the ratio of
   $v_\mathrm{Ni}^{\,\mathrm{bulk}}$ to $v_\mathrm{H}^{\,\mathrm{mix}}$:
   the greater this ratio, the smoother the distribution of hydrogen.
It turns out that the degree of overlapping is not sensitive to the explosion
   energy, as two sets of new models show in Table~\ref{tab:3Dsim}.
This reflects the fact that the smoothness of the hydrogen distribution inside
   the He shell depends mainly on the properties of the progenitor.
It is noteworthy that in our hydrodynamic models, the minimum velocity
   of free protons is lower than 400\,km\,s$^{-1}$, while hydrogen-rich matter
   mixed into the He shell expands with velocities exceeding 1200\,km\,s$^{-1}$,
   except for models B15-2 and N20-P.

Macroscopic mixing of radioactive $^{56}$Ni outward and hydrogen-rich
   matter inward continues until the ejecta reach homologous expansion.
\citet{UWJM_15} showed that the bulk of $^{56}$Ni containing $99\%$ of
   the total $^{56}$Ni mass expands almost homologously in the four old models
   B15-2, W18, W20, and N20-P already by the time when we map our 3D
   hydrodynamic data onto the 1D Lagrangian grid.
The same holds for the three new models W16-3, W18r-2, and W18x-2.
Hence, we consider our 3D neutrino-driven simulations mapped at
   $t_\mathrm{map}$, well after the phase of shock breakout at $t_\mathrm{SB}$
   (Table~\ref{tab:3Dsim}), to be acceptably close to the final results with
   respect to mixing of heavy elements in velocity space, keeping in mind that
   this extent of mixing is not exactly the final state.

\subsection{Mixing extent and properties of progenitors}
\label{sec:results-1Dpran}
%
\begin{table}[t]
\caption[]{$^{56}$Ni mixing and hydrodynamic properties of progenitors
           for explosions with similar energies.}
\label{tab:nimixing}
\centering
\begin{tabular}{@{ } l @{ } c @{ } c @{ } c @{ } c @{ } c @{ } c @{ } c @{ }}
\hline\hline
\noalign{\smallskip}
 Model & $M_\mathrm{He}^{\,\mathrm{core}}$
       & $\langle v \rangle_\mathrm{Ni}^{\mathrm{150}}$
       & \phantom{m}$v_\mathrm{Ni}^{\mathrm{150}}$
       & \phantom{m}$v_\mathrm{Ni}^{\mathrm{CO}}$
       & $\log\,(\xi/\xi_{0})^\mathrm{CO}$
       & $\Delta\,t_\mathrm{RS}$ & \phantom{m}$R_\mathrm{RS}$ \\
\noalign{\smallskip}
       & $(M_{\sun})$ & \multicolumn{3}{c}{($10^3$\,km\,s$^{-1}$)}
       &  & (s) & \phantom{m}$(R_{\sun})$ \\
\noalign{\smallskip}
\hline
\noalign{\smallskip}
 B15-2  & 4.05 & 1.20 & 2.91 & 19.85           & 6.91 & \phantom{e}612 & \phantom{m}4.21 \\
 W16-3  & 6.55 & 1.12 & 2.28 & 13.78           & 4.18 & \phantom{e}873 & \phantom{m}2.73 \\
 W18    & 7.40 & 0.85 & 1.26 & \phantom{e}8.07 & 3.75 & 1241           & \phantom{m}3.55 \\
 W18r-2 & 6.65 & 0.76 & 1.10 & \phantom{e}7.85 & 4.14 & \phantom{e}858 & \phantom{m}2.79 \\
 W18x-2 & 5.13 & 0.93 & 1.89 & 16.56           & 5.85 & \phantom{e}273 & \phantom{m}1.60 \\
 W20    & 5.79 & 0.76 & 1.18 & 12.62           & 5.54 & \phantom{e}318 & \phantom{m}1.84 \\
 N20-P  & 5.98 & 0.91 & 1.38 & \phantom{e}8.39 & 5.19 & \phantom{e}224 & \phantom{m}2.27 \\
\noalign{\smallskip}
\hline
\end{tabular}
\tablefoot{%
The two left columns give the name of the 3D hydrodynamic explosion model
   and the helium-core mass of the corresponding pre-SN model.
The next three columns provide the averaged characteristic velocities for
   the bulk of $^{56}$Ni containing $96\%$ of the total $^{56}$Ni mass in
   the 3D simulations;
$\langle v \rangle_\mathrm{Ni}^{\mathrm{150}}$ is the weighted mean velocity
   of the bulk mass of $^{56}$Ni with the $^{56}$Ni mass fraction as weight
   function at day 150;
$v_\mathrm{Ni}^{\mathrm{150}}$ is the maximum velocity of the bulk mass of
   $^{56}$Ni at the same epoch; and
$v_\mathrm{Ni}^{\mathrm{CO}}$ is the maximum velocity of the bulk mass of
   $^{56}$Ni at the moment just after the main shock has passed the (C+O)/He
   composition interface.
The last three columns give the 1D hydrodynamic properties of the progenitors:
$(\xi/\xi_{0})^\mathrm{CO}$ is the maximum time-integrated Rayleigh-Taylor
   growth factor in the close vicinity of the (C+O)/He composition interface
   (Fig.~\ref{fig:rtgrowth});
$\Delta\,t_\mathrm{RS}$ is the formation time of the reverse shock below
   the He/H composition interface, which is reckoned from the moment when the
   main shock crosses this composition interface; and
$R_\mathrm{RS}$ is the radius of the reverse shock at the formation epoch.
}
\end{table}
\citet{KPSJM_06} discovered and \citet{WMJ_15} further studied the
   sensitivity of outward $^{56}$Ni mixing and inward hydrogen mixing
   to the structure of the helium core and the He/H composition interface.
We have analyzed this sensitivity for the models B15, W18, W20, and
   N20 in a previous paper \citep{UWJM_15} and now revisit this question adding
   the new models W16, W18r, and W18x (Table~\ref{tab:presnm},
   Fig.~\ref{fig:denmr}).
To study the influence of pre-SN properties on turbulent mixing, we considered
   models B15-2, W16-3, W18, W18r-2, W18x-2, W20, and N20-P which have
   similar explosion energies (Table~\ref{tab:3Dsim}).
Along the model sequence B15-2, W16-3, W18x-2, N20-P, W18, W20, and W18r-2,
   the mixing of radioactive $^{56}$Ni monotonically decreases in velocity
   space, while the mass of hydrogen mixed into the He shell does not show
   any trend.

These correlations are of interest in understanding the global properties
   of mixing in our extended sample of BSG progenitors.
We first address the mixing of radioactive $^{56}$Ni and try to understand
   this complex phenomenon using a simple approach.
As mentioned above, there are two crucial factors that favor mixing of
   radioactive $^{56}$Ni in velocity space: (1) a fast growth of Rayleigh-Taylor
   instabilities at the (C+O)/He composition interface, and (2) a weak
   interaction of fast Rayleigh-Taylor plumes, with the strong reverse shock
   occurring below the He/H composition interface.

In order to mimic multidimensional effects of Rayleigh-Taylor mixing at the
   (C+O)/He composition interface, we considered a simple phenomenological
   approach to describe the evolution of the nickel velocity:
\begin{equation}
   {d v \over d t} = \beta \sigma_{\mathrm{RT}} v_{0} \; ,
\label{eq:model}
\end{equation}
where $v$ is the maximum velocity of the bulk mass of radioactive $^{56}$Ni,
   $\beta$ is an empirical buoyancy coefficient,
   $\sigma_{\mathrm{RT}}$ is the linear Rayleigh-Taylor growth rate
   (Eq.~\ref{eq:sigmaRT}), and
   $v_{0}$ is the initial value of the radial velocity of $^{56}$Ni.
The solution of Eq.~\ref{eq:model} is
\begin{equation}
   v(t) = \beta v_{0} \int_{0}^t \sigma_{\mathrm{RT}}(\tau) d\tau
        = \beta v_{0} \ln(\xi/\xi_0) \; ,
 \label{eq:rtsol}
\end{equation}
where $\xi/\xi_0$ is the time-integrated growth factor (Eq.~\ref{eq:rel_xi}).
Thus, the velocity growth factor is proportional to the natural logarithm of
   the time-integrated Rayleigh-Taylor growth factor.
For our consideration of the evolution of the nickel velocity, we evaluated
   the maximum time-integrated Rayleigh-Taylor growth factor in the close
   vicinity of the (C+O)/He composition interface, $(\xi/\xi_{0})^\mathrm{CO}$
   (Fig.~\ref{fig:rtgrowth}, Table~\ref{tab:nimixing}).

\begin{figure}[t]
   \includegraphics[width=\hsize, clip, trim=18 153 67 319]{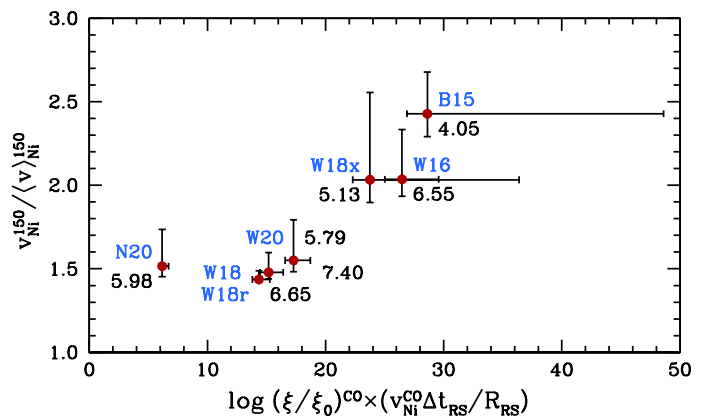}
   \caption{%
   Dependence of the extent of $^{56}$Ni mixing in velocity space on the
      hydrodynamic properties of the progenitor for models B15-2, W16-3, W18,
      W18r-2, W18x-2, W20, and N20-P (Table~\ref{tab:nimixing}),
      whose explosions have similar energies.
   The uncertainties in this $^{56}$Ni mixing are related to estimates of
      the velocities of the $^{56}$Ni ejecta containing $93\%$ and $99\%$
      of the total $^{56}$Ni mass.
   Numbers give the helium-core masses of the corresponding progenitor
      models.
   }
   \label{fig:vnipsn}
\end{figure}
How strongly the fast Rayleigh-Taylor plumes that grow from the C+O core
   through the He layer interact with the reverse shock that develops
   below the He/H composition interface depends on the ratio of the time
   interval during which the reverse shock forms relative to the time needed
   by the Rayleigh-Taylor plumes originating from the (C+O)/He interface
   to cross the reverse shock.
The greater this ratio, the weaker the interaction and the higher
   the terminal velocity of fast Rayleigh-Taylor plumes.
The formation time of the reverse shock, $\Delta\,t_\mathrm{RS}$, is estimated
   from the hydrodynamic model as the time from when the main shock crosses
   the He/H composition interface to the moment at which the reverse shock
   forms.
The characteristic time of the outward growth of Rayleigh-Taylor plumes may be
   measured by the ratio of the reverse shock radius at the formation epoch,
   $R_\mathrm{RS}$, to the characteristic velocity of the plumes.
The latter may be identified with the maximum velocity of the bulk mass of
   $^{56}$Ni at the moment just after the main shock passes the (C+O)/He
   composition interface, $v_\mathrm{Ni}^{\mathrm{CO}}$.
In Table~\ref{tab:nimixing} we give the corresponding quantities that
   determine the amount of mixing of radioactive $^{56}$Ni for models 
   B15-2, W16-3, W18, W18r-2, W18x-2, W20, and N20-P in our simple
   phenomenological approach.

Figure~\ref{fig:vnipsn} shows the extent of $^{56}$Ni mixing in velocity space
   as a function of the hydrodynamic properties for our progenitor models.
The extent of $^{56}$Ni mixing is measured by the dimensionless ratio of the
   maximum velocity of the bulk mass of $^{56}$Ni ejecta containing $96\%$ of
   the total $^{56}$Ni mass at day 150, $v_\mathrm{Ni}^{\mathrm{150}}$, to
   the weighted mean velocity of the bulk mass of $^{56}$Ni at the same epoch,
   $\langle v \rangle_\mathrm{Ni}^{\mathrm{150}}$, whereby we eliminate
   the influence of slightly different explosion energies of the models.
Because the outward mixing of $^{56}$Ni depends on the growth of Rayleigh-Taylor
   plumes at the (C+O)/He composition interface as well as on the interaction
   of those plumes with the reverse shock from the He/H composition interface,
   we use the product of these two physically independent measures as
   abscissa in Fig.~\ref{fig:vnipsn}.

Obviously, there is a correlation between the normalized extent of
   $^{56}$Ni mixing and the two hydrodynamic properties that favor the mixing of
   radioactive $^{56}$Ni in different progenitor models (Fig.~\ref{fig:vnipsn}).
The existence of this correlation confirms the decisive role of the two
   selected factors on the amount of outward mixing of radioactive $^{56}$Ni
   in the framework of 3D neutrino-driven simulations.
In addition, the model sequence B15-2, W18x-2, W20, N20-P, W16-3, W18r-2, and
   W18 shows that strong radioactive $^{56}$Ni mixing tends to be disfavored
   by a high helium-core mass of the progenitor model.
The latter result is very important when constructing adequate progenitor
   models that might produce an amount of mixing of radioactive $^{56}$Ni
   compatible with the spectroscopic observations of SN~1987A.

We note that this correlation becomes more obvious when we exclude model N20-P
   from our consideration (Fig.~\ref{fig:vnipsn}).
Similarly, this model is also inconsistent with the decreasing trend of the
   amount of hydrogen mixed into the He shell in the model sequence B15-2,
   W16-3, W18x-2, N20-P, W18, W20, and W18r-2, along which the mixing of
   radioactive $^{56}$Ni monotonically decreases in velocity space.
Thus, the progenitor model N20 differs considerably from the other models in
   producing macroscopic mixing of both radioactive $^{56}$Ni and hydrogen.
It might be that the outlier status of model N20 is related to its
   nonevolutionary origin, as pointed out in Sect.~\ref{sec:modmeth-psn}.

We do not place much weight on model N20 for these reasons.
The present set of BSG explosion models exhibits $^{56}$Ni mixing whose
   strength scales roughly inversely with the helium-core mass
   $M_\mathrm{He}^{\,\mathrm{core}}$ (see the numbers next to the model names
   in Fig.~\ref{fig:vnipsn}).
This may appear plausible because the growth factor
   $\log(\xi/\xi_0)^\mathrm{CO}$ decreases nearly monotonically with
   $M_\mathrm{He}^{\,\mathrm{core}}$ (Table~\ref{tab:nimixing}), which might
   suggest that $v_\mathrm{Ni}^{\mathrm{CO}}$ simply grows with higher growth
   factor (i.e., lower helium-core mass), thus leading to $^{56}$Ni penetration
   of the hydrogen envelope with higher velocities.
However, models W18 and W18r-2 do not perfectly follow this ordering in
   Fig.~\ref{fig:vnipsn}, and in particular, model W16-3 constitutes a clear
   case that contradicts such a simple causal relation: W16-3 exhibits
   a fairly high value of $v_\mathrm{Ni}^{\mathrm{CO}}$ despite its rather
   low value of $\log(\xi/\xi_0)^\mathrm{CO}$ (see Table~\ref{tab:nimixing}).

Nevertheless, we do not consider model W16-3 as a peculiar outlier with
   respect to its value of $v_\mathrm{Ni}^{\mathrm{CO}}$, which requires
   that some special physics play a role in this particular model.
A detailed analysis of the explosion properties and evolution did not reveal
   any peculiarities in this case.
Instead, we interpret the situation such that there is \emph{no clear} trend
   of $v_\mathrm{Ni}^{\mathrm{CO}}$ growing with $\log(\xi/\xi_0)^\mathrm{CO}$,
   but the results of $v_\mathrm{Ni}^{\mathrm{CO}}$ show basically random
   scatter as function of the growth factor: models W16-3, W18, and W18r-2
   have very similar values of $\log(\xi/\xi_0)^\mathrm{CO}$, but their 
   $v_\mathrm{Ni}^{\mathrm{CO}}$ values differ considerably, whereas 
   models N20-P, W20, W18x-2, and B15-2 possess significantly larger growth
   factors (more than an order of magnitude, increasing for the model sequence
   given) than model W16-3, but model N20-P has a lower and model W20 has
   a similar value of $v_\mathrm{Ni}^{\mathrm{CO}}$ compared to model W16-3.
Therefore we conclude that the growth factor at the (C+O)/He composition
   interface alone \emph{cannot} explain the final $^{56}$Ni-mixing velocities.
Instead, our measure of the $^{56}$Ni mixing on the vertical axis of
   Fig.~\ref{fig:vnipsn} shows a tight correlation with the product of
   the two factors coining the quantity on the horizontal axis of this figure.
The second factor, whose physical meaning is explained above by the effects
   of an interaction of the $^{56}$Ni plumes with the reverse shock from
   the He/H composition interface, moves model W20 with its low value of 
   $v_\mathrm{Ni}^\mathrm{150}/\langle v \rangle_\mathrm{Ni}^\mathrm{150}$
   to the left, while model W16-3 stays more toward the right, although
   both models have nearly the same values of $v_\mathrm{Ni}^{\mathrm{CO}}$.

Of course, all this reasoning is based on grounds of a fairly limited sample
   of 3D simulations.
A simple ordering of the $^{56}$Ni mixing with the growth factor at
   the (C+O)/He composition interface, $\log(\xi/\xi_0)^\mathrm{CO}$,
   (or, equivalently, an inverse correlation with the helium-core mass)
   mostly contradicts models W16-3 and W20.
To conclusively clarify whether both of these models are special cases
   requires computing a considerably larger set of BSG explosions
   with similar energies.

\subsection{Light-curve modeling}
\label{sec:results-lcmod}
%
\begin{figure}[t]
   \includegraphics[width=\hsize, clip, trim=18 153 75 321]{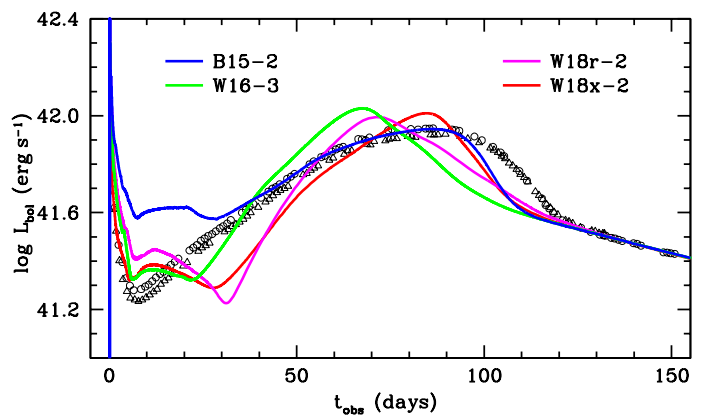}
   \caption{%
   Bolometric light curves of models B15-2, W16-3, W18r-2, and W18x-2 compared
      with the observations of SN~1987A obtained by \citet{CMM_87, CWF_88}
      (open circles) and \citet{HSGM_88} (open triangles). 
   }
   \label{fig:lcobs}
\end{figure}
We \citep{UWJM_15} have in detail performed light-curve modeling
   based on 3D hydrodynamic simulations of neutrino-driven explosions for
   the set of pre-SN models B15, W18, W20, and N20.
In particular, we studied the influence of the explosion energy and the
   ejected $^{56}$Ni mass on the calculated light curves.
Here, we focus on the properties of the new pre-SN models W16, W18r, and W18.

Our previous hydrodynamic light-curve modeling showed that the pre-SN radii of
   all models considered are too large to fit the initial luminosity peak of
   SN~1987A.
Our hydrodynamic simulations with models W16, W18r, and W18x,
   whose radii are smaller than those used previously (Table~\ref{tab:presnm},
   Fig.~\ref{fig:denmr}b), show a much better agreement of the
   calculated light curves with the observed initial luminosity peak than
   reference model B15-2 (Fig.~\ref{fig:lcobs}).
In addition, models W16-3, W18r-2, and W18x-2 are able to synthesize amounts
   of radioactive $^{56}$Ni in the general range of what guarantees a good match of
   the calculated light curves with the observed radioactive tail
   (Table~\ref{tab:3Dsim}, Fig.~\ref{fig:lcobs}).

For SN~1987A, the total mass of radioactive $^{56}$Ni, evaluated by equating
   the observed bolometric luminosity in the radioactive tail to the gamma-ray
   luminosity and called the observed amount of radioactive $^{56}$Ni,
   is $0.0722\,M_{\sun}$ \citep{UWJM_15}.
The $^{56}$Ni mass in the ejecta at day 150, $M^{f}_{\mathrm{Ni}}$, which
   matches the observed luminosity in the radioactive tail, exceeds the
   observed value because of the expansion-work effects \citep{Utr_07}.
The initial $^{56}$Ni mass at the onset of light curve modeling,
   $M^{i}_{\mathrm{Ni}}$, exceeds in turn the value of $M^{f}_{\mathrm{Ni}}$
   because of fallback of $^{56}$Ni taken  into account.
The initial $^{56}$Ni masses of models W16-3, W18r-2, and W18x-2 fall
   in between the minimum, $M_\mathrm{Ni}^{\,\mathrm{min}}$, and maximum,
   $M_\mathrm{Ni}^{\,\mathrm{max}}$, values (Table~\ref{tab:3Dsim}).

While the initial luminosity peak mainly depends on the pre-SN radius and the
   structure of the outer layers, the radioactive tail is entirely
   determined by the total amount of radioactive $^{56}$Ni.
The light curve in between these phases depends in addition on the explosion
   energy, the ejecta mass, and, in what we are interested here, the
   macroscopic mixing of $^{56}$Ni and hydrogen-rich matter.
A comparison of the light curves of models W16-3, W18r-2, and W18x-2 with
   that of model B15-2 shows that there are significant differences in
   the rise to the second maximum of the light curve and its dome-like look.
The quality of the light-curve fits can be only partially improved by
   enhancing the macroscopic mixing of $^{56}$Ni and hydrogen-rich matter.
The differences in the dome-shaped part of the light curve also reflect
   to some extent the differences in the pre-SN structure.

Of particular interest is the question of the role of 3D macroscopic mixing in
   light-curve modeling of ordinary versus peculiar type IIP SN.
We discuss this in Appendix~\ref{sec:apndx}.

\subsection{Comparison with observations}
\label{sec:results-cmpobs}
%
\begin{figure}[t]
   \includegraphics[width=\hsize, clip, trim=18 153 75 321]{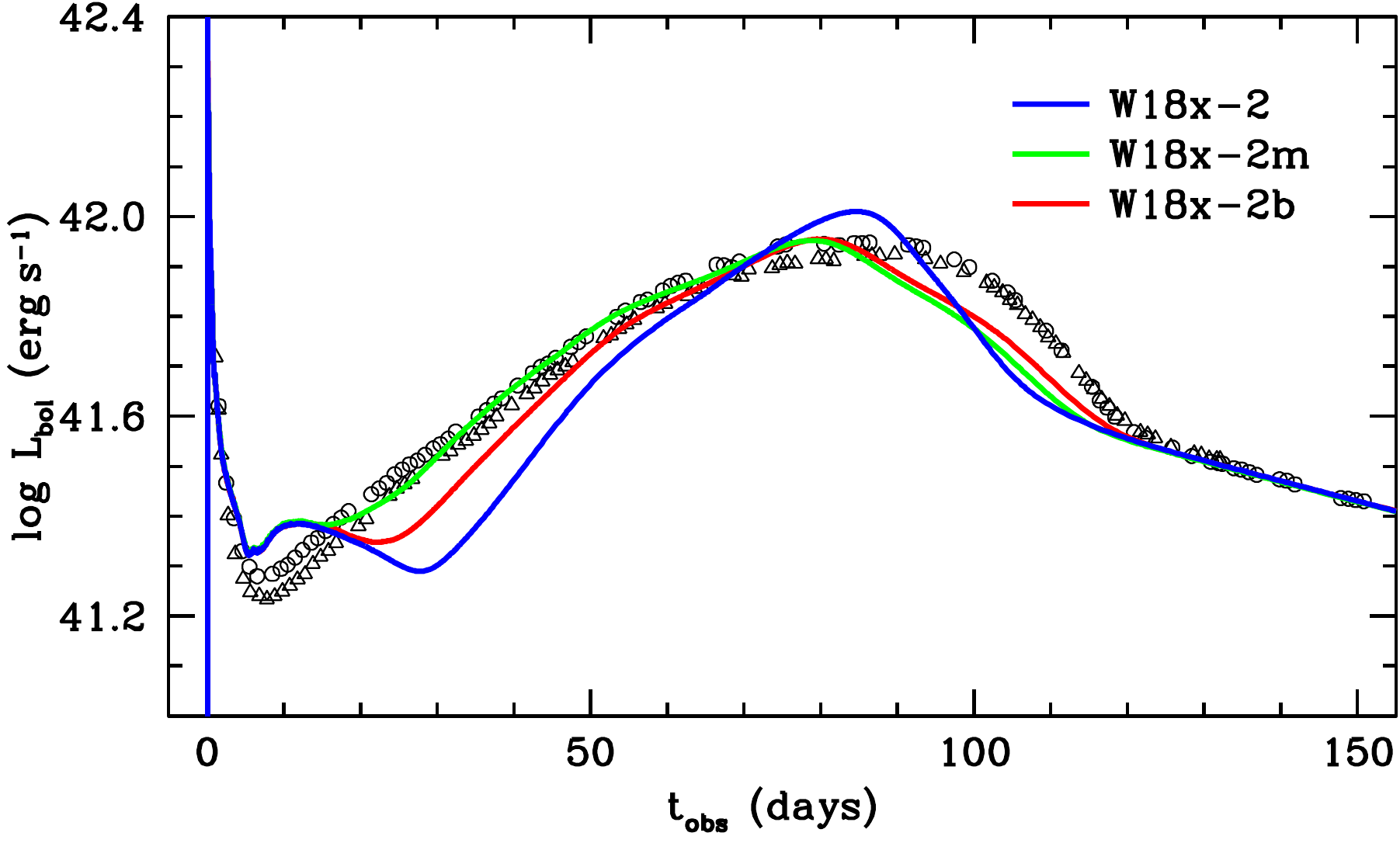}
   \caption{%
   Bolometric light curves of model W18x-2 and two additional models W18x-2m
      and W18x-2b compared with the observations of SN~1987A obtained by
      \citet{CMM_87, CWF_88} (open circles) and \citet{HSGM_88}
      (open triangles).
   Model W18x-2m is constructed on the basis of model W18x-2, where both
      $^{56}$Ni and hydrogen are mixed artificially to match the observations
      as well as possible.
   Outward $^{56}$Ni mixing up to 3500\,km\,s$^{-1}$ and inward H mixing of
      0.26\,$M_{\sun}$ into the He core result in a good fit.
   Model W18x-2b is similar to model W18x-2m, but the bulk of $^{56}$Ni is
      mixed up to the observed velocity of about 3000\,km\,s$^{-1}$.
   The two additional models W18x-2m and W18x-2b have the unchanged initial
      $^{56}$Ni mass of model W18x-2.
   }
   \label{fig:lcfit}
\end{figure}
The light curve computed by \citet{UWJM_15} for the reference model B15-2
   fits the dome-shaped part of the observed light curve of SN~1987A during
   the rise to its maximum much better than those of the other previous
   models N20-P, W18, and W20.
Model B15-2 also yields a maximum velocity of the bulk of $^{56}$Ni
   that is consistent with the observed value of about 3000\,km\,s$^{-1}$.
However, a very disappointing result was that the calculated light curves of
   all these models, which have the SN~1987A explosion energy, disagree
   with the observations during the first $30-40$ days.
The new models W16-3, W18r-2, and W18x-2, which have smaller pre-SN radii
   than those of the old models (Table~\ref{tab:presnm}), demonstrate
   a much better agreement of the calculated light curves with the observations
   during this early period, in particular, for the initial luminosity peak
   (Fig.~\ref{fig:lcobs}).

Of the new models, W18x-2 has the most acceptable dome-shape of the light
   curve, in particular, with respect to the position of the main maximum.
We therefore focus now on this model.
A closer look reveals two essential disparities between the calculated and
   observed light curves for this model: a pronounced local luminosity minimum
   at day 27, and a too narrow, more peak-like main light-curve maximum.
With the pre-SN structure being specified, these shortcomings can be improved
   by enhancing the macroscopic mixing.
First, the more intense the macroscopic mixing of $^{56}$Ni in velocity
   space, the earlier the luminosity starts to grow again to the broad   
   dome-shaped maximum.
As a consequence, the local minimum at around day 27 becomes more shallow.
Second, inward mixing of hydrogen-rich matter increases the opacity and
   optical depth of the inner ejecta, which in turn give rises to a wider and
   less luminous dome-shaped light-curve maximum.

The resulting effect of enhanced artificial macroscopic mixing is
   illustrated by the additional model W18x-2m in Fig.~\ref{fig:lcfit}.
Imposing an outward mixing of $^{56}$Ni up to 3500\,km\,s$^{-1}$ and
   an inward mixing of 0.26\,$M_{\sun}$ of hydrogen into the helium core
   in the hydrodynamic light-curve modeling, we obtain a better agreement
   with the observations of SN~1987A than for model W18x-2.
The agreement is only partial, and two shortcomings related to the pre-SN
   properties still remain.
The small bump in the computed luminosity at day 10 is caused by the inadequate
   density distribution in the outer layers.
Reducing the relatively strong $^{56}$Ni mixing up to a velocity of
   3500\,km\,s$^{-1}$ in model W18x-2m to the observational constraint of
   about 3000\,km\,s$^{-1}$ in the additional model W18x-2b causes the bump
   to become more prominent and emphasizes the shortcoming of the pre-SN structure.
The luminosity deficit during the transition from the broad maximum to
   the radioactive tail can be mitigated by an increase in ejecta mass and
   a proportional increase in explosion energy \citep{UWJM_15}.
   
In the sequence of models W18r-2, W20, W18, N20-P, W18x-2, W16-3, and B15-2,
   with comparable and SN~1987A-like explosion energies and a growing amount
   of mixing of radioactive $^{56}$Ni in velocity space (Table~\ref{tab:3Dsim},
   Fig.~\ref{fig:vnipsn}),
   only model B15-2 yields the maximum velocity of the bulk mass of $^{56}$Ni
   of 3370\,km\,s$^{-1}$ consistent with the spectral observations of SN~1987A,
   while all other models fall short of the required mixing.
However, model B15, with the lowest helium-core mass of 4.05\,$M_{\sun}$
   of the available pre-SN models and the right amount of $^{56}$Ni
   mixing, is inconsistent with the observational data of Sanduleak
   $-69^{\circ}202$, which suggest a single star with a helium-core mass
   of about 6\,$M_{\sun}$.

The analysis of the line profiles in the nebular phase \citep{Chu_91, KF_98, MJS_12}
   and the 3D view of molecular hydrogen in SN~1987A \citep{LSF_19} showed that
   hydrogen is mixed into the core to velocities $\le$700\,km\,s$^{-1}$.
Hydrodynamic models based on 3D neutrino-driven simulations with SN~1987A-like
   explosion energies demonstrate that hydrogen in the innermost
   layers of the ejecta expands with velocities lower than 100\,km\,s$^{-1}$,
   in good agreement with the observations.
In addition, \citet{KF_98} constrained the mass of hydrogen-rich gas mixed
   within 2000\,km\,s$^{-1}$ to about 2.2\,$M_{\sun}$.
Our favorite models B15-2 and W18x-2 yield hydrogen matter mixed within
   2000\,km\,s$^{-1}$ of 0.922\,$M_{\sun}$ and 0.847\,$M_{\sun}$, respectively
   (Table~\ref{tab:3Dsim}), which is in qualitative agreement with the
   observational constraint.

\section{Conclusions}
\label{sec:conclsn}
%
Performing 3D neutrino-driven explosion simulations and subsequent light curve
   modeling, we compared the results with the observations of SN~1987A and
   draw the following main conclusions:
\begin{itemize}
\item
Hydrodynamic light-curve modeling for a set of previously used pre-SN models
   (B15, W18, W20, and N20) showed that the pre-SN radii of all these models
   are too large to fit the initial luminosity peak of SN~1987A.
In addition, the structure of the outer layers of these models cannot reproduce
   the observed light curve during the first 30--40 days \citep{UWJM_15}.
Hydrodynamic simulations with new progenitor models (W16, W18r, and W18x)
   that have smaller radii than the other models demonstrate a much better
   agreement of the calculated light curves with the observed one in the
   initial luminosity peak and during the first 20 days than our old
   hydrodynamic models.
\item
The initial (at the onset of light-curve modeling) $^{56}$Ni masses required
   to match the observations of the radioactive tail of SN~1987A fall in
   between the minimum and maximum estimates obtained in our 3D explosion
   models, implying that all 3D neutrino-driven simulations under
   study are able to synthesize the ejected amount of radioactive $^{56}$Ni
   for explosion energies in the general range of what is needed to explain the
   observed light curve.
\item
A sequence of models (B15-2, W18x-2, W20, N20-P, W16-3, W18r-2, and W18) with
   nearly the same explosion energies showed that the extent of outward
   radioactive $^{56}$Ni mixing in the framework of the 3D neutrino-driven
   simulations depends mainly on the following two hydrodynamic properties
   of the progenitor model:
A high growth factor of Rayleigh-Taylor instabilities at the (C+O)/He
   composition interface, and a weak interaction of fast Rayleigh-Taylor
   plumes, with the reverse shock occurring below the He/H composition
   interface, seem to be a sufficient condition for efficient outward mixing
   of $^{56}$Ni into the hydrogen envelope.
\item
The analysis of SN~1987A observations revealed the fact that radioactive
   $^{56}$Ni was mixed in the ejecta up to a velocity of about
   3000\,km\,s$^{-1}$.
In our 3D neutrino-driven simulations, only model B15-2 yields a maximum
   velocity of the bulk of $^{56}$Ni consistent with the observations,
   and only this model is therefore able to reproduce the increase to
   the light-curve maximum of SN~1987A with good agreement.
However, the helium-core mass of $4.05\,M_{\sun}$ of the pre-SN model
   B15 is inconsistent, assuming single-star evolution, with the 
   observational data of the BSG Sanduleak $-69^{\circ}202$ star, which 
   suggest a star that has a helium-core mass of $\approx$6\,$M_{\sun}$
   at the time of its explosion.
\end{itemize}

From the 3D simulations we presented for macroscopic mixing that occurs during
   neutrino-driven explosions, we conclude that existing evolutionary models
   of single stars cannot simultaneously fulfill the observational requirements
   of the location of the BSG Sanduleak $-69^{\circ}202$ star in the
   Hertzsprung-Russell diagram and the observed amount and extent of outward
   mixing of radioactive $^{56}$Ni.
Two speculative possibilities for a solution of this problem have been proposed: 
   a progenitor with a more rapidly rotating iron core and a jet-induced
   explosion \citep{KHO_99, WMW_02, WWH_02}, or a binary evolution scenario
   \citep{HM_89, PJ_89, PJR_90, PJH_92}.
While the question of rapid rotation of the iron core remains open, the
   formation of the SN~1987A triple-ring system was explained by
   \citet{PMI_07} and \citet{MP_09} in the scenario of a binary merger model.

In our study the rotating progenitor models W16, W18, W18r, and W18x have
   moderate angular momenta on the main sequence of
   $(2.7, 3.1, 3.1, \mathrm{and}\ 2.3)\times10^{52}$\,erg\,s, respectively
   (Sect.~\ref{sec:modmeth-psn}).
To estimate an initial rotation rate of the resultant pulsar, we assumed that
   the inner 1.7\,$M_{\sun}$ core collapses to a newly born neutron star
   with a gravitational mass of 1.4\,$M_{\sun}$ and a radius of 12\,km, and
   that angular momentum is conserved during the collapse.
For models W16, W18, W18r, and W18x, the estimates for the spin periods
   (or, to be more accurate, their lower limits) are 10.6, 10.3, 12.5,
   and 10.0\,ms, respectively.
The transport of angular momentum by magnetic fields during the collapse
   might have increased the period, but we note that even this maximum value
   would imply a small contribution from rotation to the explosion energy.
We ignored such a progenitor rotation because it is so slow that it does not
   affect the neutrino-driven mechanism, nor is it an efficient
   energy source for a magnetohydrodynamics mechanism.
This is consistent with the lack of any bright pulsar in the current
   remnant \citep{ALF_18}, which supports slow rotation of the degenerate
   stellar core.
Therefore a scenario involving the formation of (jittering) jets
   \citep{Sok_17, BS_18} seems disfavored as well.

However, slow core rotation does not preclude that rotation, for instance, from
   a binary progenitor scenario, contributes to mixing and asymmetry in the
   pre-collapse star and in the explosion.
\citet{MH_17} presented the results of a stellar evolution study of binary
   merger models for the progenitor of SN~1987A.
Their Hertzsprung-Russell diagram shows that binary merger models with
   a primary star of 16\,$M_{\sun}$ and secondaries ranging from 
   2 to 8\,$M_{\sun}$ evolve to compact pre-SN models in or close to the
   region of the observed properties of Sanduleak $-69^{\circ}202$.
The helium-core masses of all of these pre-SN models are lower than 4\,$M_{\sun}$,
   which seems promising for producing the high-velocity $^{56}$Ni-rich ejecta
   observed in the spectra of SN~1987A.
Recently, \citet{MUH_19} have made the first successful step on the road of
   applying these binary merger progenitors to hydrodynamically modeling
   the light curve of SN~1987A, which was computed with an artificial ``boxcar'' mixing
   of the chemical species and an explosion by a piston in spherically
   symmetric geometry.
The appropriate 3D simulations of neutrino-driven explosions of these SN~1987A
   progenitor models are in progress to determine the corresponding mixing
   self-consistency.

We conclude that the lack of an adequate progenitor model for the well-observed
   and well-studied SN~1987A remains a challenge for the theory of stellar
   evolution of massive stars.

\begin{acknowledgements}
%
V.P.U. was supported by the guest program of the Max-Planck-Institut f\"ur
   Astrophysik.
S.E.W. acknowledges support by the NASA Theory Program (NNX14AH34G).
At Garching, funding by the Deutsche Forschungsgemeinschaft through grant
   EXC 153 ``Origin and Structure of the Universe'' and by the European
   Research Council through ERC-AdG No.\ 341157-COCO2CASA is acknowledged.
The 3D models were computed and the data were postprocessed on Hydra
   of the Rechenzentrum Garching.
%
\end{acknowledgements}



\begin{appendix}
%
\section{Role of 3D macroscopic mixing in light-curve modeling of
         ordinary and peculiar type IIP supernovae}
\label{sec:apndx}
%
\begin{figure*}[t]
\centering
   \includegraphics[width=0.48\hsize, clip, trim=18 153 66 321]{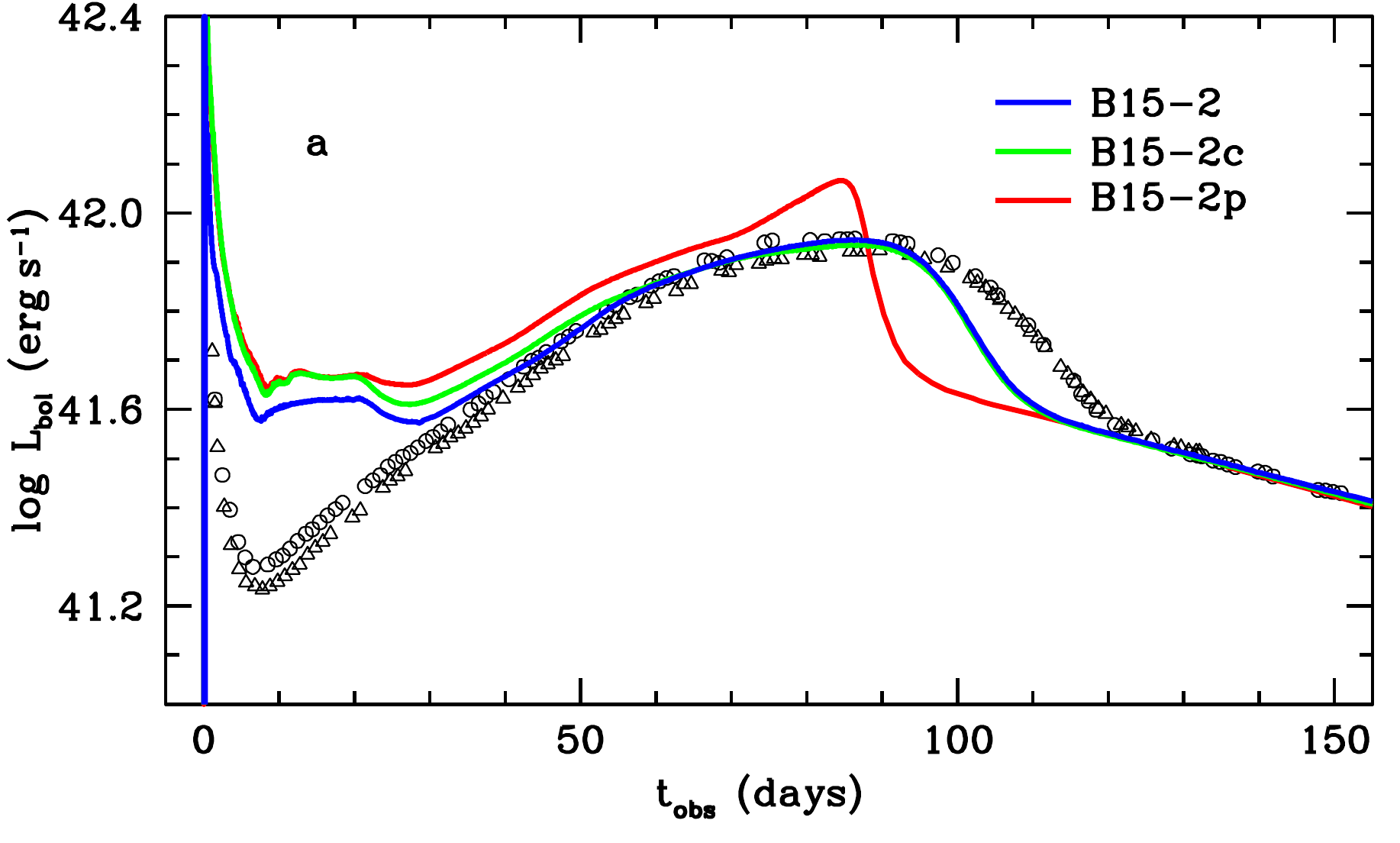}
   \hspace{0.5cm}
   \includegraphics[width=0.48\hsize, clip, trim=18 153 66 321]{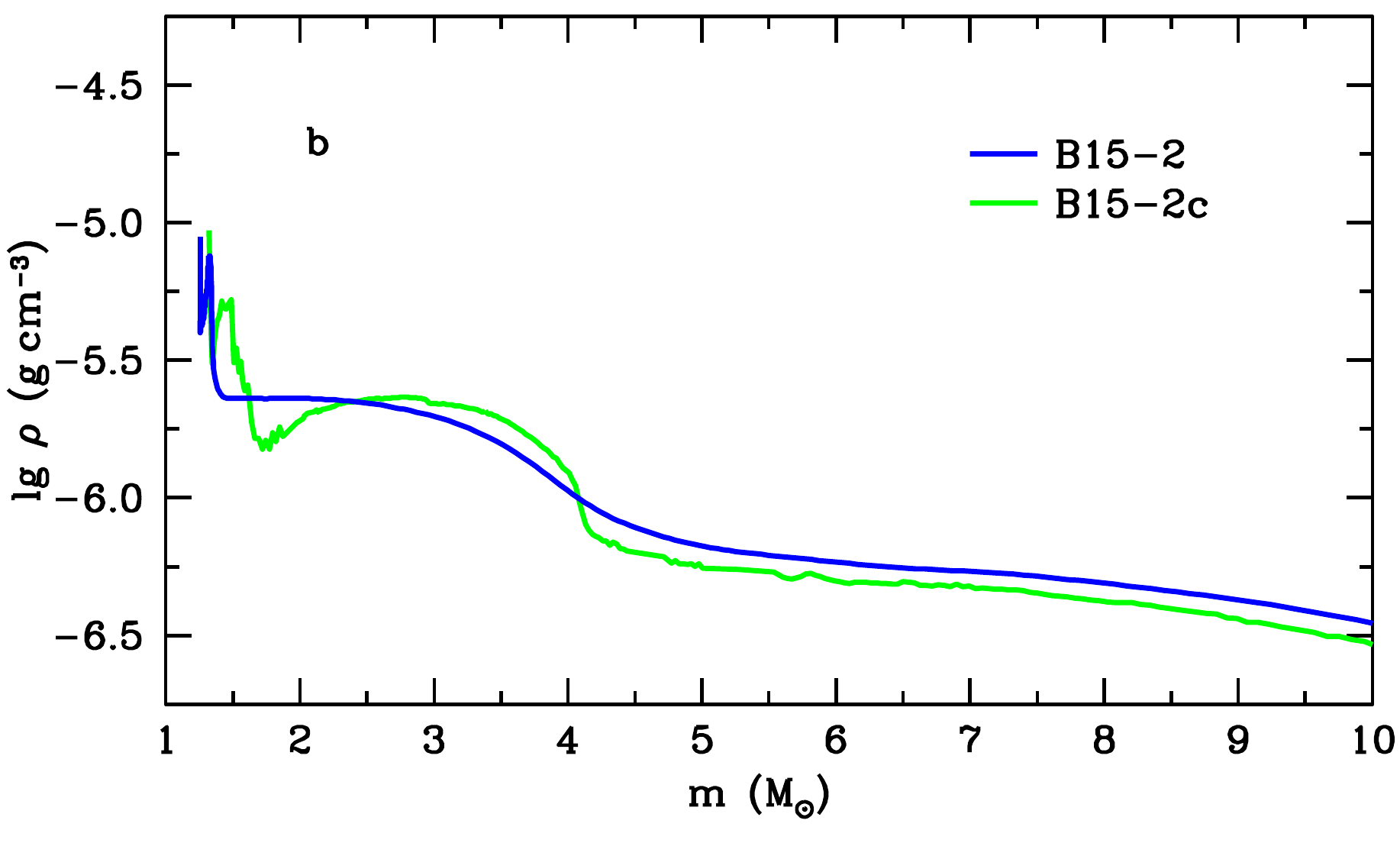}\\
   \caption{%
   Effect of 3D macroscopic mixing on the bolometric light curve (a) and
      the density profile as a function of interior mass at the mapping moment
      of $t=61,213$\,s (b).
   The reference model B15-2 (blue) is compared with model B15-2c (green),
      which is the 1D analog of model B15-2.
   Model B15-2c is based on the pre-SN model B15, is exploded by a piston,
      and its chemical composition is identical to that of the averaged
      3D model B15-2.
   For a comparison, the red line shows the bolometric light curve of model
      B15-2p, which is based on the unmixed pre-SN model B15 and exploded by
      a piston.
   Open circles \citep{CMM_87, CWF_88} and open triangles \citep{HSGM_88}
      are the bolometric data of SN~1987A.
   }
   \label{fig:3dvs1d}
\end{figure*}
The first attempts to hydrodynamically model the peculiar type IIP SN~1987A
   faced an unexpected problem: the original, unmixed evolutionary pre-SN
   models produced a half-truncated maximum of the light curve like that of
   model B15-2p in Fig.~\ref{fig:3dvs1d}a \citep[cf.][]{Woo_88, AF_89}.
This result is inconsistent with the observations of SN~1987A.
A similar problem arose in modeling ordinary type IIP SNe with
   evolutionary pre-SN models when theoretical light curves exhibited
   a conspicuous shoulder-like \citep{WH_07, MPR_15, SEWBJ_16} or spike-like
   \citep{CDHLS_03, You_04} feature during the decline in luminosity from
   the plateau to the radioactive tail, which is not seen in observations.
For both peculiar and ordinary type IIP SNe, artificial
   mixing of the chemical species was invoked to solve these problems.

The study of the ordinary type IIP SN~1999em showed that turbulent mixing
   induced by 3D neutrino-driven explosions causes both flattening of the
   density distribution at the location of the He/H interface and macroscopic
   mixing of chemical species, which in turn cause the monotonic
   decline in luminosity from the plateau to the radioactive tail as it is
   observed \citep{UWJM_17}.
In our previous paper on the peculiar type IIP SN~1987A \citep{UWJM_15}, we
   focused mainly on macroscopic mixing of radioactive $^{56}$Ni and
   hydrogen that occur during the explosion.
Here we revisit the issue of turbulent mixing in SN~1987A in the context of
   modifying the density distribution compared to 1D hydrodynamic models.

To clarify the influence of mixing on the density distribution and the light
   curve for the explosion of a BSG star, we compared our reference model B15-2
   (Table~\ref{tab:3Dsim}) with its 1D analog B15-2c.
The latter is based on the pre-SN model B15 (Table~\ref{tab:presnm}), is
   a piston-driven explosion, and its chemical composition is identical
   to that of the averaged 3D model B15-2 (Fig.~\ref{fig:mfvel}a).
Model B15-2c thus constructed permits us to refine the effect of turbulent
   mixing on the density distribution at the location of the He/H composition
   interface from other effects.

Figure~\ref{fig:3dvs1d}b shows that a density step (contact discontinuity)
   at the outer edge of the helium core (at about 4\,$M_{\sun}$) appears
   after the SN shock crossed the He/H composition interface in 1D model
   B15-2c, while the averaged density distribution of the 3D model B15-2 is
   smooth and has no signature of a density step at the location of the He/H
   interface, similar to the 3D explosion of a RSG star \citep{UWJM_17}.
The smoothness of the density distribution reflects a characteristic feature
   of our 3D neutrino-driven explosion simulations: large-scale
   turbulent mixing occurs, which smoothes the density distribution.
The less prominent density step in the case of model B15-2c compared
   to that of model L15-pn \citep[][see Fig.~8a]{UWJM_17} is related to
   a weaker density contrast at the outer edge of the helium core in the BSG
   model B15 (Fig.~\ref{fig:denmr}c) compared to that of the RSG model L15
   \citep[][see Fig.~2a]{UWJM_17}.

The density structure within the helium core of the 1D hydrodynamic
   model B15-2c does not affect the shape of the light curve, which is similar
   to that of the reference model B15-2 (Fig.~\ref{fig:3dvs1d}a).
The slightly higher luminosity of model B15-2c compared to that of
   model B15-2 during the first 20 days is caused by the difference in density
   outside the helium core (Fig.~\ref{fig:3dvs1d}b).
The lower density in model B15-2c means that the star is more extended
   (a given mass is distributed in a larger volume), leading to larger
   radiation losses.
This situation raises the question why the 1D density structure for
   the peculiar type IIP SN~1987A does not exhibit any conspicuous spike-like
   feature that is not observed, whereas it does in the 1D hydrodynamic model
   for the ordinary type IIP SN~1999em \citep{UWJM_17}.
There are two reasons behind this behavior.
First, the step-like feature in density at the He/H interface in model B15-2c
   is less prominent than that of model L15-pn, as pointed out above.
Second, the photosphere crosses the sharp He/H interface in model B15-2c at
   around day 85 (Fig.~\ref{fig:3dvs1d}a) and in model L15-pn at around
   day 135 \citep[][see Fig.~7]{UWJM_17}, when the sources of the internal
   energy in these models are quite different.

At around day 85, the luminosity of the peculiar type IIP SN~1987A is
   completely powered by gamma-rays from radioactive $^{56}$Co, while the
   internal energy deposited by the shock wave has already been exhausted
   by about day 30.
The difference in density between models B15-2c and B15-2 in
   the mass range from 2 to 4\,$M_{\sun}$ (Fig.~\ref{fig:3dvs1d}b)
   does not cause an essential energy excess deposited by gamma-rays from
   radioactive $^{56}$Co in model B15-2c compared to model B15-2, and as a
   consequence, results in a close similarity between the corresponding
   light curves near maximum (Fig.~\ref{fig:3dvs1d}a).
At the same time, a comparison of model B15-2p based on the unmixed pre-SN
   model B15 with model B15-2c, which has a chemical composition identical to
   that of the averaged 3D model B15-2, shows that mixing induced by the 3D
   neutrino-driven explosion eliminates the unobserved half-truncated maximum
   of the light curve of model B15-2p (Fig.~\ref{fig:3dvs1d}a).
Thus, turbulent mixing that occurs during the 3D neutrino-driven explosion of
   a BSG star causes both smoothing of the density distribution and macroscopic
   mixing of the chemical species, but only the latter affects the light
   curve.

In contrast to SN~1987A, the release of the internal energy deposited during
   the propagation of the shock wave through the pre-SN envelope produces
   the luminosity of the ordinary type IIP SN~1999em not only during the
   plateau phase, but also during the monotonic decline in luminosity from the
   plateau to the radioactive tail.
In turn, the release of the internal energy deposited by radioactive decays
   becomes essential just before entering the radioactive tail.
At around day 135 in model L15-pn for SN~1999em, when a pronounced spike
   appears in the luminosity decline from the plateau to the radioactive tail,
   the internal energy deposited by the shock remains dominant in powering
   the luminosity.
\citet{UWJM_17} showed that both the density step at the outer edge of the
   helium core and the unmixed chemical composition of the evolutionary pre-SN
   model are responsible for the unobserved spike in the light curve that is
   computed with hydrodynamic models that are exploded by a piston in spherically
   symmetric geometry, and that this spike disappears only in the framework of
   the 3D neutrino-driven explosion simulations.
%
\end{appendix}

\end{document}